\documentclass[aip,jcp,reprint]{revtex4-2}

\usepackage[svgnames]{xcolor}
\usepackage{amsmath}
\usepackage{bm}
\usepackage{braket}
\usepackage{listings}
\usepackage{amssymb}
\usepackage{graphicx}
\usepackage{comment}
\graphicspath{{./img/}}

\begin{document}

\title{Open quantum dynamics theory for a complex subenvironment system with a quantum thermostat: Application to a spin heat bath}

\author{Kiyoto Nakamura}
\email{nakamura.kiyoto.n20@kyoto-u.jp}
\author{Yoshitaka Tanimura}
\email{tanimura.yoshitaka.5w@kyoto-u.jp}
\affiliation{Department of Chemistry, Graduate School of Science, Kyoto University, Sakyoku, Kyoto 606-8502, Japan}

\date{\today}

\begin{abstract}
Complex environments, such as molecular matrices and biological material, play a fundamental role in many important dynamic processes in condensed phases. Because it is extremely difficult to conduct full quantum dynamics simulations on such environments due to their many degrees of freedom, here we treat in detail the environment only around the main system of interest (the subenvironment), while the other degrees of freedom needed to maintain the equilibrium temperature are described by a simple harmonic bath, which we call a quantum thermostat. The noise generated by the subenvironment is spatially non-local and non-Gaussian and cannot be characterized by the fluctuation-dissipation theorem. We describe this model by simulating the dynamics of a two-level system (TLS) that interacts with a subenvironment consisting of a one-dimensional $XXZ$ spin chain. The hierarchical Schr\"odinger equations of motion are employed to describe the quantum thermostat, allowing time-irreversible simulations of the dynamics at arbitrary temperature.
To see the effects of a quantum phase transition of the subenvironment, we investigate the decoherence and relaxation processes of the TLS at zero and finite temperatures for various values of the spin anisotropy. We observed the decoherence of the TLS at finite temperature, even when the anisotropy of the $XXZ$ model is enormous. We also found that the population relaxation dynamics of the TLS changed in a complex manner with the change of the anisotropy and the ferromagnetic or antiferromagnetic orders of the spins.
\end{abstract}

\maketitle


\section{INTRODUCTION}
Ensembles of molecular systems with competing interactions have long been an important topic in physics, chemistry, and biology. Such materials exhibit various intriguing behaviors. For example, in a spin glass system or an amorphous solid, longtime relaxation processes play a key role as the temperature falls, leading to a slowing down of the dynamic response. In superconducting qubits, the spin environment exhibits peculiar non-Ohmic effects.\cite{shnirman05sb,Weiss2012}
As noise sources, these materials also have unique properties. In standard open quantum dynamics theories, thermal noise is assumed to be Gaussian and uniformly distributed in space. The noise is assumed to come from a heat bath described by an infinite set of harmonic oscillators at finite temperature.\cite{Breuer2002,Weiss2012,caldeira83cl,CLModel,LeggettRMP87A} The heat bath is considered to be an unlimited heat source with infinite heat capacity.\cite{ST20JCP, ST21JSPS} However, the thermal noise from complex or frustrated materials is typically non-Gaussian, and its temporal and spatial correlations are non-uniform, although the ensemble average of the noise has a Gaussian profile when the strength of each source of noise is comparable, due to the central limit theorem.\cite{vankampen2007spp} The noise correlation is characterized by a simple function, for example, a stretched exponential function.

These peculiar features of noise in complex environments have been observed in single molecular spectroscopy using impurity molecules as a probe.\cite{Moerner1991,Skiner1995,Skiner1996,TTK98JCP} The spectral random walk of the transition energies of each molecule can be separately measured. Such noise that depends on the location of the environment can also be measured by muon spin spectroscopy, which is used to investigate  spin glass, superconductivity, and biological materials. \cite{Uemura1980,Matsuda,Takeshita2009,Hiraishi2014,TORIKAI2006441,TT20JPSJ} The possibility of characterizing a spin bath through the decoherence of a single qubit has also been suggested.\cite{cucchietti05sb, cucchietti07sb}

Although problems of this kind are well established and are experimentally measurable, theoretical investigations, in particular in a quantum regime, are challenging due to the complexity of the environment. Although dynamics simulations have been conducted to investigate the energy fluctuations of the environment, they have been limited to classical or phenomenological cases using a molecular dynamics simulation,\cite{Kramer2014,LeeCoker2016,UT21JCTC} a stochastic approach,\cite{Skiner1995} or the Metropolis Monte Carlo approach.\cite{TTK98JCP} If the environment is an ensemble of harmonic oscillators, such approaches are useful for characterizing the noise because the response function of the noise is identical in both classical and quantum cases. However, such characterization does not work for non-Gaussian and non-local quantum noise, because the fluctuation--dissipation theorem (FDT) does not hold. As a result, conventional open quantum dynamics theory is unable to elucidate the effects of complex environments.

Because it is extremely difficult to conduct simulations of quantum dynamics for complex environments with many degrees of freedom and where the heat capacity of the environment is required to be infinity, the practical approach is to treat a part of the environment around the main system of interest as a subenvironment. The thermodynamic effects of the other degrees of freedom are described using a simple harmonic bath, which we call a quantum thermostat (QT). To use a heat bath as the thermostat, we use weak subenvironment--bath coupling with a simple spectral distribution function to suppress the dynamic aspects of the bath while maintaining the temperature of the subenvironment. When there are sufficiently many subenvironmental degrees of freedom and if we keep the coupling strength of the thermostat weak, we should be able to study the effects of a complex environment in a quantum-mechanically consistent manner. 
We employ the hierarchical equations of motion (HEOM) to describe the thermostat, assuming the relation is based on the FDT for symmetric and antisymmetric correlation functions of the noise operators, which is equivalent to assuming the harmonic heat bath is coupled to the subenvironment.\cite{TK89JPSJ1,T90PRA,T06JPSJ,T14JCP,T15JCP,T20JCP}

To demonstrate our approach, we employ a two-level system (TLS) that interacts with a one-dimensional (1D) spin lattice. Such systems have been investigated to enhance the performance of quantum information processing.\cite{nielsen00}  Moreover, molecular magnet systems\cite{leuenberger01mm, dobrovitski00mm} and Josephson junction systems\cite{kenyon00jj,shnirman05jj} are described by this model.
In the past, various forms of a 1D spin lattice, such as the Ising,\cite{venuti10z, haikka12z} $XY$,\cite{yuan07xy, cormick08xy, cheng09xy, hu10xy, wu14xx} and $XXZ$\cite{lai08xxz, rossini07sb, yuan08xxx} models, have been investigated. 
Such spin-lattice systems, when isolated, exhibit quantum phase transitions depending on the system parameters.\cite{franchini17}
For the Ising and $XY$ lattices, the spins are ordered (i.e., the magnetization is finite) when the transverse magnetic field is small, while the spins are disordered when the magnetic field becomes large.
For an $XXZ$ lattice, the spins have ferromagnetic and antiferromagnetic phases depending on the anisotropy.  Between these quantum phases, there is a critical region where the excitation spectrum is gapless.\cite{franchini17}
The effects of such quantum phase transitions and critical regions on a TLS have been extensively studied.

Here we investigate a spin-lattice system, such as a heat bath, using the QT described by the wavefunction-based hierarchical Schr{\"o}dinger equations of motion (HSEOM).\cite{NT18PRA} 
By controlling the temperature of the thermostat, we investigate the thermodynamic aspects of a spin-lattice system in a practical way. 

This paper is organized as follows. In Sec.~\ref{sec:model}, we discuss the QT described by the HSEOM. Then in Sec.~\ref{sec:spinbath}, we introduce a subenviroment described by the 1D $XXZ$ chain. 
The results of the numerical simulations of a TLS interacting with a 1D spin-lattice environment are discussed in Sec.~\ref{sec:results}.
Section~\ref{sec:conclude} has our concluding remarks.

\section{Quantum thermostat} \label{sec:model}

\subsection{System--subenvironment model coupled to a QT}

We consider a main system (S) coupled to a subenvironment (SE) expressed as 
\begin{align}
\hat{H}_{\mathrm{S+SE}} = \hat{H}_{S} + \hat{H}_{S-SE} + \hat{H}_{SE},
\label{eq:SpinH}
\end{align}
where $\hat{H}_{S}$, $\hat{H}_{S-SE}$, and $\hat{H}_{SE}$ are the Hamiltonians of the system, system--subenvironment interaction, and subenvironment, respectively. 
To take into account the time irreversibility and temperature effects of the subenvironment, we introduce a QT described by the Gaussian quantum noise operator $\hat \Omega(t)$.
The total Hamiltonian is then expressed as:
\begin{align}
\hat{H}_{\mathrm{S+SE}}'(t) = \hat{H}_{\mathrm{S+SE}} - \hat {V}_{\mathrm{SE}}{\hat \Omega}(t),
\label{eq:stochastic} 
\end{align}
where $\hat {V}_{\mathrm{SE}}$ is the subenvironment part of the noise interaction. 
For bosonic noise, ${\hat \Omega}(t)$ is characterized by the antisymmetric and symmetric correlation functions that satisfy Kubo's FDT as  
\begin{align}
\langle [\hat \Omega (t), \hat \Omega(0)]\rangle_{\Omega}=-2i\hbar\int_0^{\infty}  d\omega J(\omega)\sin(\omega t)
\label{eq:CorrIm}
\end{align}
and 
\begin{align}
\frac{1}{2}\langle \hat \Omega(t) \hat \Omega(0)+\hat \Omega(0) \hat \Omega(t) \rangle_{\Omega}=& \nonumber \\
\hbar \int_{0}^{\infty}  d\omega &J(\omega) \coth\left( \frac{\beta\hbar\omega}{2}\right) \cos(\omega t),
\label{eq:CorrRe}
\end{align}
where $J(\omega)$ is the spectral distribution function (SDF), $\beta \equiv 1/k_{\mathrm{B}}T$ is the inverse temperature divided by the Boltzmann constant $k_\mathrm{B}$, and $\langle \dots{} \rangle_{\Omega}$ represents the expectation values of the noise trajectories. The antisymmetric correlation function that describes the dissipation of the SE system is independent of temperature, whereas the symmetric correlation function that describes the fluctuations depends on the temperature. They are related through the FDT. The total system approaches thermal equilibrium in the longtime limit.\cite{TK89JPSJ1,T06JPSJ} Note that because of the function $\coth(\beta\hbar\omega/2)$, the fluctuations cannot be Markovian, even if we choose a \textit{strictly Ohmic} SDF, $J(\omega) = \gamma \omega$.\cite{Weiss2012} Thus, a non-Markov treatment of the noise is critical for the quantum thermal state, especially for low temperatures.\cite{T06JPSJ,T14JCP,T15JCP,T20JCP}

By taking the ensemble average with respect to ${\hat \Omega}(t)$, the system described by Eqs.~\eqref{eq:SpinH} and~\eqref{eq:stochastic} is equivalent to the SE system coupled to the harmonic heat bath (or the QT) described as:
\begin{align}
\hat{H}_{\mathrm{QT}} =  \sum\limits_{j} {\left( \frac{\hat p_j^2 }{2m_j } + \frac{1}{2}m_j \omega _j^2 \hat x_j^2 \right) } - \hat V_{\mathrm{SE}} \sum\limits_j {c_j \hat x_j}.
\label{eq:totH2}
\end{align}
The noise in the second term on the right-hand side in Eq.~\eqref{eq:stochastic}, ${\hat \Omega}(t)$, corresponds to the term  $\sum_j {c_j \hat x_j}$ in Eq.~\eqref{eq:totH2} with the SDF defined as $J(\omega) \equiv \sum_{j} c_{j}^2/(2m_{j}\omega_{j}) \delta(\omega-\omega_{j})$.

There are two possible approaches for investigating the time evolution of an S+SE system coupled to a QT. The first approach handles the noise explicitly.  We define the noise operator using a sequence of stochastic operators, $\hat \Omega (t)=\{ \hat \Omega_1(t), \hat \Omega_2(t), \hat \Omega_3(t),\dots \}$, which satisfy the relations Eqs.~\eqref{eq:CorrIm} and~\eqref{eq:CorrRe}.\cite{T06JPSJ}  Then, for different noise trajectories $\hat \Omega_j(t)$, we numerically integrate the quantum Liouville equation, $d\hat{\rho}_{\mathrm{S+SE}}(t; \hat \Omega_j(t))/dt = i[\hat{H}_{\mathrm{S+SE}}'(t), \hat{\rho}_{\mathrm{S+SE}}(t; \hat \Omega_j(t))]/\hbar$, many times to accumulate the stochastic samples.  The quantum dynamics of the S+SE system are then investigated by averaging over the samples, as $\hat{\rho}_{\mathrm{S+SE}}(t) \propto \sum_{j} \hat{\rho}_{\mathrm{S+SE}}(t; \hat \Omega_j(t))$.
At a glance, this approach looks similar to the classical (or semi-classical) Langevin approach used in simulations of Brownian particles.
In the Langevin approach, the equation contains a friction term in addition to a random modulation term, and the friction term is related to the random modulation term through the FDT. The present equation of motion, however, does not involve a friction term. Friction appears through the antisymmetric correlation function of the noise operator. Thus, to describe the effects of dissipation, the noise is often defined using complex variables. \cite{Stockburger1999,Stockburger2001,Stockburger2002,Shao2004} Note that the treatment of the quantum noise operator is not easy, especially at low temperatures, as the symmetric correlation function has a negative value shorter than the time scale of the inverse Matsubara frequency. \cite{T06JPSJ,T20JCP}
Thus, various stochastic equations of motion have been derived in which a hierarchical structure is employed for the dissipation part, while the fluctuation part is treated as noise. \cite{CaoStochastic2013, Shao2004,Yun-anShao2004}

In the second approach, instead of utilizing a sequence of noise operators, we use the equation of motion for the averaged density operator, $\hat{\rho}_{\mathrm{S+SE}}(t)=\langle\hat{\rho}_{\mathrm{S+SE}}(t; \hat \Omega(t))\rangle_{\Omega}$.  For a bosonic Gaussian bath, this equation of motion is the same as the reduced HEOM.\cite{TK89JPSJ1,T90PRA,T06JPSJ,T14JCP,T15JCP,T20JCP} The relation between the first approach and the second approach is similar to the relation between the Langevin equation approach and the Fokker--Planck equation approach in the classical case.

While the approaches based on the explicit treatment of the noise and those based on the averaged density operator are identical as long as the noise is bosonic Gaussian, the first has wider applicability. For example, if we replace $\coth(\beta\hbar\omega/2)$ with 
$\tanh(\beta\hbar\omega/2)$, we can generate fermionic noise, although the argument becomes phenomenological. Extensions of the hierarchical stochastic approach to non-Gaussian quantum noise have been proposed, \cite{CaoStochastic2018A,CaoStochastic2018B} in which higher-order correlation (or cumulant) functions to determine the property of the noise are utilized.
 
\subsection{Hierarchical Schr{\"o}dinger equations of motion}

In the present approach, it is necessary to treat large systems with many degrees of freedom, because the subenvironment, which consists of numerous quantum states, is explicitly treated as $\hat{H}_{\mathrm{S+SE}}$.
In a specific case, we could treat a spin bath by mapping it to a bosonic bath using an effective noise correlation function,\cite{CaoStochastic2018B} but this is not possible in the current situation.
In the conventional open quantum dynamics theory for a system with $N$ states, an $N \times N$ reduced density matrix has to be used for a time-irreversible process described by a non-Hermitian propagator. However, the memory required to calculate the density matrix elements is enormous. 
Here, we adopt the HSEOM,\cite{NT18PRA} whose scalability is similar to that of the Schr\"odinger equation. In this formalism, the time evolution of the left-hand side (ket vector) is computed from time $0$ to $t$, and subsequently, the time evolution of the right-hand side (bra vector) is computed from time $t$ to $0$ along the contour in the complex time plane. While conventional HEOM approaches for the reduced density operators use a set of exponential functions to describe the bath correlation function, the HSEOM use orthogonal functions to maintain the stability of the equations of motion when moving along the contour in the direction of decreasing time. When a special function is chosen as a basis, while calculations at zero temperature become easier, simulations for long times become difficult due to the increase in the number of bases.\cite{ZhoufeiWu2015,WuCao2017}
By appropriately choosing functions for the bath correlation function and the spectral density, we can take advantage of an HSEOM method to conduct simulations without needing a huge amount of computational memory.

Such wavefunction-based approaches have also been developed based on an explicit treatment of noise, including the stochastic hierarchy of pure states,\cite{Strunz2014,Strunz2015,Strunz2017A,Strunz2017B,Strunz21sto} the stochastic Schr\"odinger equation,\cite{Shi2016Sto} the stochastic Schr\"odinger equation with a diagonalized influence functional along the contour in the complex time plane, \cite{cao96sto} and a hierarchy of stochastic Schr\"odinger equations. \cite{KeZhao2016,KeZhao2017,KeZhao2017B,KeZhao2018,KeZhao2019,WangKeZhao2019} Although the formulation of the stochastic approaches is considerably simpler than the HSEOM approach, it is not suitable for studying a system with slow relaxation or a system subjected to a slowly varying time-dependent external force, because the convergence of the trajectories is slow in the stochastic approaches. For this reason, here we use the HSEOM approach. 

Technically, the HSEOM are not regarded as equations of motion, because they are not defined with the time $t$ but with the complex time contour $0 \rightarrow t \rightarrow 0$. Thus, in the HSEOM approach, we must repeat the full integration for $t$ along $0 \rightarrow t+\Delta t \rightarrow 0$ for different $\Delta t $, which is in contrast to the conventional HEOM approach. We remove this difficulty by applying the method developed by Xu \textit{et al.},\cite{xu11heomheis} which was originally developed to calculate higher-order time-correlation functions.
The set of HSEOM we solve here are expressed as (Appendix A):
\begin{align}
\frac{\partial}{\partial t}  \ket{\phi_{\vec{n}} (t)} = &
- \frac{i}{\hbar} \hat{H}_{\mathrm{S+SE}} \ket{\phi_{\vec{n}} (t) } \nonumber \\
&+ \sum_{k=0}^{K-1} \sum_{k'=0}^{K-1} \eta_{k, k'} n_{k}
\ket{ \phi_{\vec{n} - \vec{e}_{k} + \vec{e}_{k'}}  (t)} \nonumber \\
& - \frac{i}{\hbar} \hat{V}_{\mathrm{SE}} \sum_{k=0}^{K-1} c_{k} \ket{  \phi_{\vec{n} + \vec{e}_{k}}  (t)} \nonumber \\
& - \frac{i}{\hbar} \hat{V}_{\mathrm{SE}}  \sum_{k=0}^{K-1} n_{k} \varphi_{k} (0)
\ket{\phi_{\vec{n} - \vec{e}_{k}}  (t)} ,
\label{eq:SEOMnewA}
\end{align}
\begin{align}
\frac{\partial}{\partial t}  \ket{\psi_{\vec{n}} (t)} = &
- \frac{i}{\hbar} \hat{H}_{\mathrm{S+SE}} \ket{\psi_{\vec{n}} (t) } \nonumber \\
& + \sum_{k=0}^{K-1} \sum_{k'=0}^{K-1} \eta_{k, k'} (n_{k}+1)
\ket{ \psi_{\vec{n} + \vec{e}_{k} - \vec{e}_{k'}}  (t)} \nonumber \\
& - \frac{i}{\hbar} \hat{V}_{\mathrm{SE}}  \sum_{k=0}^{K-1} c_{k}^{*} \ket{  \psi_{\vec{n} - \vec{e}_{k}}  (t)} \nonumber \\
& - \frac{i}{\hbar} \hat{V}_{\mathrm{SE}}  \sum_{k=0}^{K-1} (n_{k}+1) \varphi_{k} (0)
\ket{\psi_{\vec{n} + \vec{e}_{k}}  (t)}.
\label{eq:SEOMnewB}
\end{align}
Here, the vector $\vec{n} = [n_{0}, \ldots, n_{K-1}]$, which consists of non-negative integers, is introduced to describe the non-Markovian effects through the auxiliary wavefunctions (AWFs), $\ket{\phi_{\vec{n}} (t)}$ and $\ket{\psi_{\vec{n}} (t)}$.
The vector, $\vec{e}_{k}$, is the unit vector in the $k$th direction.

The density operator for the total spins (the main system and the subenviroment) is defined as:
\begin{align}
\hat{\rho}_{\mathrm{S+SE}}(t) = \sum_{\vec{n}} \ket{\phi_{\vec{n}}(t)} \bra{\psi_{\vec{n}}(t)}.
\end{align}
The set $\{c_k\}$ in Eqs.~\eqref{eq:SEOMnewA} and~\eqref{eq:SEOMnewB} contains the expansion coefficients that describe the two-time correlation function of the bosonic Gaussian bath, and $K$ is the number of coefficients.
We use Bessel functions to approximate the two-time correlation function of the noise and evaluate $\{c_{k}\}$ using the Jacobi--Anger identity.\cite{Guanhua12chev, Kleine16chev}
The set $\{\eta_{k, k'}\}$ contains the derivative coefficients of the Bessel functions, $d J_k (t) / dt = \sum_{k'}\eta_{k, k'} J_{k'}(t)$.

We considered an Ohmic SDF with a circular cutoff: \cite{Ando98, Guanhua12chev}
\begin{align}
J(\omega) = \zeta \omega\sqrt{1 - (\omega/\nu)^{2}},
\label{eq:circular}
\end{align}
where $\zeta$ and $\nu$ are the coupling strength and cutoff frequency, respectively. It has been reported that the numerical results obtained with this cutoff exhibit similar behavior to the Ohmic case.
If we choose $\nu = \gamma$ and $\zeta = 2 \eta / e$ (where $e$ is the base of the natural logarithm), under the condition $\omega_{0} \ll \gamma$, where $\omega_{0}$ is the characteristic frequency of the system, then the SDF can be expressed as $J(\omega) = \eta \omega e^{-|\omega|/\gamma}$.\cite{NT18PRA}

Note that similar hierarchical Schr{\"o}dinger-type equations have been developed. The bath bosons were mapped in an appropriate form, and the two-time reduced density matrices were calculated rigorously.\cite{tokieda20}

\section{Spin-lattice subenvironment} \label{sec:spinbath}

To demonstrate the capability of our approach, here we consider a TLS described as:
\begin{align}
\hat{H}_{S} = -\frac{\hbar}{2} \omega_{0} \hat{\sigma}_{0}^{z},
\label{eq:tls}
\end{align}
where $\omega_{0}$ is the excitation energy of the TLS for the eigenstate $\ket{\pm}$. The subenvironment of the 1D spin lattice is expressed as:
\begin{align}
\hat{H}_{SE} = - \frac{\hbar}{2} J\sum_{j=1}^{N-1} \left(\hat{\sigma}_{j}^{x} \hat{\sigma}_{j+1}^{x}
+\hat{\sigma}_{j}^{y} \hat{\sigma}_{j+1}^{y} + \Delta \hat{\sigma}_{j}^{z} \hat{\sigma}_{j+1}^{z} \right),
\label{eq:xxz}
\end{align}
where $N$, $J$, and $\Delta$ are the number of spins, the interaction strength between neighboring spins, and the anisotropy, respectively. The operators $\hat{\sigma}^{\alpha}_{j} \, (\alpha \in \{x, y, z\})$ are the Pauli matrices, $j=0$ represents the TLS, and $j \in [1, N]$ represents the $j$th spin in the spin lattice.

The $XXZ$ model can be in one of various quantum phases.\cite{franchini17}
When the anisotropy $\Delta>1$, it is in the ferromagnetic phase, whereas it is in the antiferromagnetic phase when $\Delta < -1$.
Between those states ($-1 \leq \Delta \leq 1$), the excitation spectrum becomes gapless and the lattice system is in the critical region.
Although a large range of models have been considered for $\hat{H}_{SE}$, we adopt the $XXZ$ model in this paper with the open boundary condition.

For the interaction between the TLS and the spins in the spin lattice, $\hat{H}_{S-SE}$, we consider (a) diagonal interactions (pure dephasing):
\begin{align}
\hat{H}_{S-SE} = -\frac{\hbar}{2} \epsilon_0 \hat{\sigma}_{0}^{z} \sum _{\mu = 1} ^{M} \hat{\sigma}_{j_\mu}^{z}
\label{eq:diag}
\end{align}
and (b) off-diagonal interactions: 
\begin{align}
\hat{H}_{S-SE} = -\frac{\hbar}{2} \epsilon_0
\left(\hat{\sigma}_{0}^{x} \sum_{\mu = 1}^{M} \hat{\sigma}_{j_\mu}^{x} + \hat{\sigma}_{0}^{y} \sum_{\mu = 1}^{M} \hat{\sigma}_{j_\mu}^{y}\right).
\end{align}
Here, the number of spins interacting with the TLS is denoted as $M$ ($1 \le M \le N$), and $\epsilon_0$ is the coupling strength between the TLS and the spins. 
In case (a), $\hat{H}_{S}$ and $\hat{H}_{S-SE}$ are commutable, and therefore, the population of the TLS, $\braket{+|\mathrm{tr}_{SE} \{\hat{\rho}_{\mathrm{S+SE}}(t)\} |+}$ and $ \braket{-|\mathrm{tr}_{SE} \{\hat{\rho}_{\mathrm{S+SE}}(t)\}|-}$, does not change over time, whereas the off-diagonal elements, $\braket{+|\mathrm{tr}_{SE} \{\hat{\rho}_{\mathrm{S+SE}}(t) \} |-}$ and $ \braket{-|\mathrm{tr}_{SE} \{\hat{\rho}_{\mathrm{S+SE}}(t)\}|+}$, become decoherent.
Here, $\mathrm{tr}_{SE} \{~\}$ is the partial trace of the subenviroment, and $\hat{\rho}_{\mathrm{S+SE}}(t) = e^{-i\hat{H}_{\mathrm{S+SE}} t/\hbar} \hat{\rho}_{\mathrm{S+SE}}(0) e^{iH_{\mathrm{S+SE}} t/\hbar}$ is the density operator at time $t$.
In case (b), through the off-diagonal interactions between the TLS and the spin lattice, thermal excitation and population relaxation occur, in addition to the decoherence of TLS. 

We consider that all of the spins of the subenviroment are uniformly coupled to the QT as:
\begin{align}
\hat{V}_{\mathrm{SE}} =  \hbar \sum_{k=1}^{N} (\hat{\sigma}_{k}^{x} + \hat{\sigma}_{k}^{y}).
\end{align}

\section{NUMERICAL RESULTS} \label{sec:results}

For the initial state of the simulation, we use the thermal equilibrium state of the subenviroment coupled to the QT without the TLS--spin interactions. 
If the coupling between the spin lattice and the QT is weak, the distribution approaches the Boltzmann distribution of the spin lattice itself.
For the numerical calculations, we diagonalize the Hamiltonian of the spin lattice as $\hat{H}_{SE} \ket{n;\alpha_n}=\varepsilon_{n} \ket{n;\alpha_n}$, where ${\alpha_n}$ is the label for the degenerate eigenstates whose eigenenergy is $\varepsilon_n$.  Then, the initial state in the zero-temperature case is expressed as $\hat{\rho}_{\mathrm{S+SE}}(0) = (\ket{+} + \ket{-})(\bra{+} + \bra{-}) / 2 \otimes \sum_{\alpha_0} \ket{0;\alpha_0}\bra{0;\alpha_0} / \sum_{\alpha_0}1$, whereas that in the finite inverse temperature case (finite $\beta$) is expressed as $\hat{\rho}_{\mathrm{S+SE}}(0) = (\ket{+} + \ket{-})(\bra{+} + \bra{-}) / 2 \otimes \sum_{n} \left(\sum_{\alpha_n} e^{-\beta \varepsilon_{n}} \ket{n;\alpha_n}\bra{n;\alpha_n} \right) /Z_{\mathrm{S+SE}}$, where $Z_{\mathrm{S+SE}}=\sum_{n} \left(\sum_{\alpha_n} e^{-\beta \varepsilon_{n}}\right)$.

To evaluate the density operator in the HSEOM approach, we have to compute each $\alpha_n$ element
of the density operator at time $t$ from pure initial state $\hat{\rho}_{\mathrm{S+SE}}^{n, \alpha_n}(0) = (\ket{+} + \ket{-})(\bra{+} + \bra{-}) / 2 \otimes \ket{n; \alpha_n} \bra{n; \alpha_n}/Z_{\mathrm{S+SE}}$ and then obtain $\hat{\rho}_{\mathrm{S+SE}}(t) = \sum_{n} \left(\sum_{\alpha_n} \hat{\rho}_{\mathrm{S+SE}}^{n, \alpha_n}(t) \right)$.  For zero temperature, we can obtain $\hat{\rho}_{\mathrm{S+SE}}(t)$ by repeating the evaluation of $\hat{\rho}_{\mathrm{S+SE}}^{0, \alpha_0}(t)$  for different $\alpha_0$.  The number of repeated calculations is the same as the degeneracy of the ground state and 2 for $\Delta = \pm 2$ and $-1$, and 14 for $\Delta = 1$ for the spin lattice with $N = 13$. For finite temperature, the number of states $\alpha_n$ for $n>0$ included in the calculation of $\hat{\rho}_{\mathrm{S+SE}}(t)$ increases dramatically (up to $2^{13}$ for $N=13$), which makes the computational cost very expensive.  We, thus, reduce the number of eigenstates to be considered using $Z_{\mathrm{SE}} = \mathrm{tr} \{e^{-\beta \hat{H}_{\mathrm{SE}}}\}$, which keeps $99 \%$ of its absolute value.  For $\beta \hbar \omega_0 = 2$, the number of states reduces from $2^{13}$ to $24$, $90$, $1260$, and $150$ for $\Delta = -2$, $-1$, $1$, and $2$, respectively. Note that this reduction in the number of eigenstates applies only to the initial states of the density operator $\hat{\rho}_{\mathrm{S+SE}}(t)$.  We consider all the eigenstates in the following calculations. 

Because we evaluate the ket and bra vectors of the density operator separately, there are numerical errors in the commutators and anticommutators for the conventional HEOM. 
Due to these errors, the total population of spin states $\mathrm{tr} \{\hat{\rho}_{\mathrm{S+SE}}(t)\} = \sum_{\vec{n}} \braket{\psi_{\vec{n}}| \phi_{\vec{n}}}$ is slightly smaller than 1. This error can be suppressed by normalizing the calculated density operator as $\hat{\rho}'_{\mathrm{S+SE}}(t) = \hat{\rho}_{\mathrm{S+SE}}(t) / \mathrm{tr}\{\hat{\rho}_{\mathrm{S+SE}}(t)\}$.  Hereafter, we denote $\hat{\rho}'_{\mathrm{S+SE}}(t)$ as $\hat{\rho}_{\mathrm{S+SE}}(t)$.

We chose $\omega_{0}$ as the unit for the frequency and set $\epsilon_0/\omega_{0} = J/\omega_{0} = 1$.
The number of spins in the spin lattice $N$ is set to $13$ or $15$. 
The coupling strength of the thermostat is set to a weak value of $\hbar \zeta = 0.01$, and the cutoff frequency is set to $\nu / \omega_{0} = 2$.
In the numerical calculation, we truncate the AWFs of the HSEOM with the condition $\sum_{k} n_k > 2$. For zero temperature, $\beta \hbar \omega_0 \to \infty$. For low temperatures, $\beta \hbar \omega_0 = 2$. We simulate the dynamics up to the times $t\omega_0 = 50$ and $t \omega_0 = 20$, respectively. 
The number of Bessel functions are $100$ for $\beta \hbar \omega_0 \to \infty$ and $40$ for $\beta \hbar \omega_0 = 2$.

In previous studies, all the spins of the lattice were coupled to the TLS ($M$ is set to $N$) with uniform coupling strength\cite{haikka12z, yuan07xy, cormick08xy, cheng09xy, hu10xy, wu14xx, lai08xxz} or site-dependent coupling strength.\cite{yuan08xxx}
The latter reflects the actual coupling strength of the TLS--lattice interactions that depend on the location of the TLS in the spin lattice.\cite{Skiner1995,Skiner1996,TTK98JCP}  
In this paper, we assume that the TLS--spin interactions are highly localized, and we set $M = 1$ and $j_{1} = (N+1)/2$ (the TLS is coupled only to the center spin of the spin lattice for odd $N$).\cite{rossini07sb}

\subsubsection{Characterizing the subenvironmental noise} \label{sec:corr}

\begin{figure}
\centering
\includegraphics[width=\linewidth]{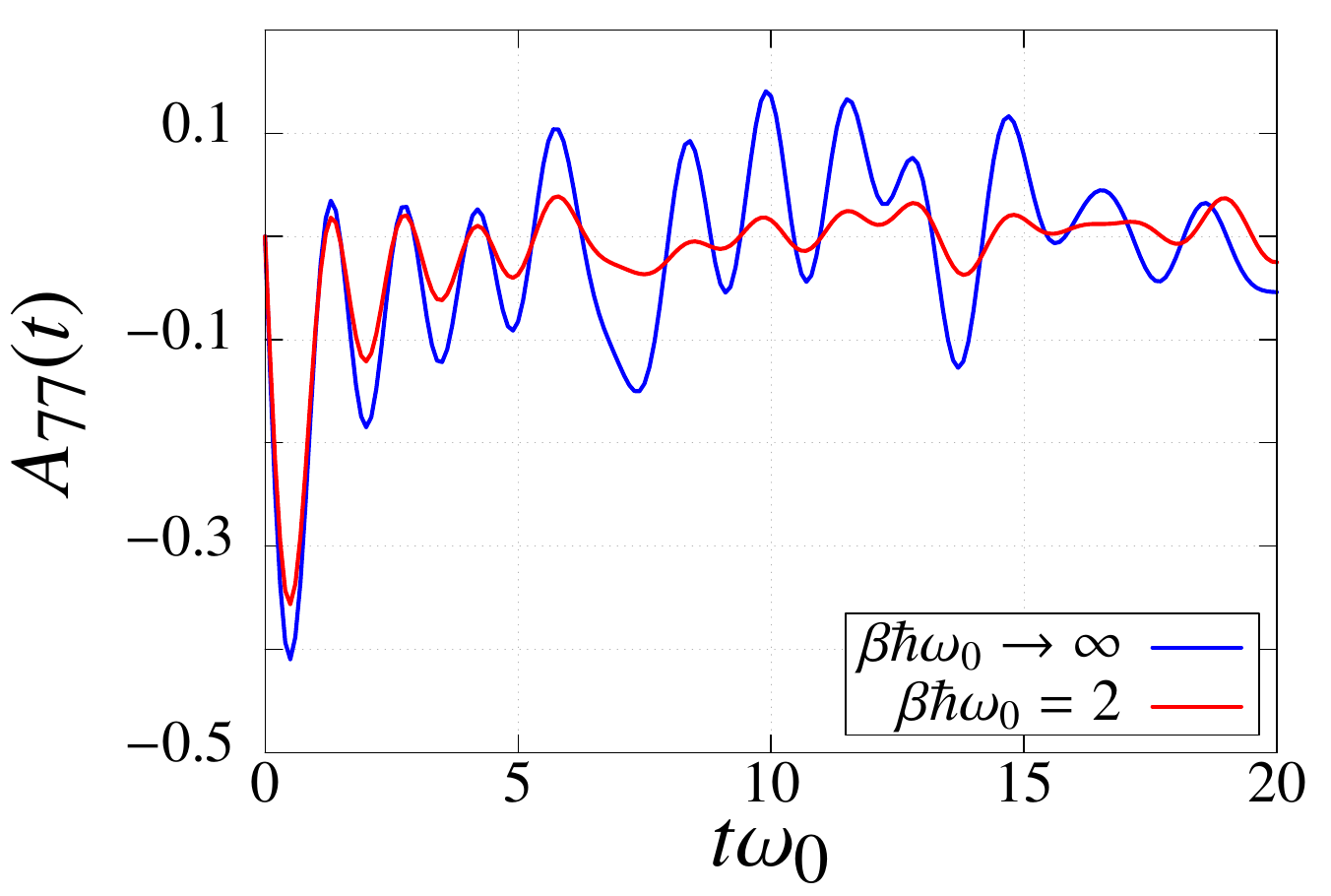}
\caption{
Antisymmetric correlation function $A_{77}(t)$ for the central spin in the 1D $XXZ$ spin lattice in arbitrary units at zero temperature ($\beta \hbar \omega_0 \to \infty$) and finite temperature ($\beta \hbar \omega_0 = 2$).
\label{fig:2corrTemp}
}
\end{figure}

\begin{figure}
\centering
\includegraphics[width=\linewidth]{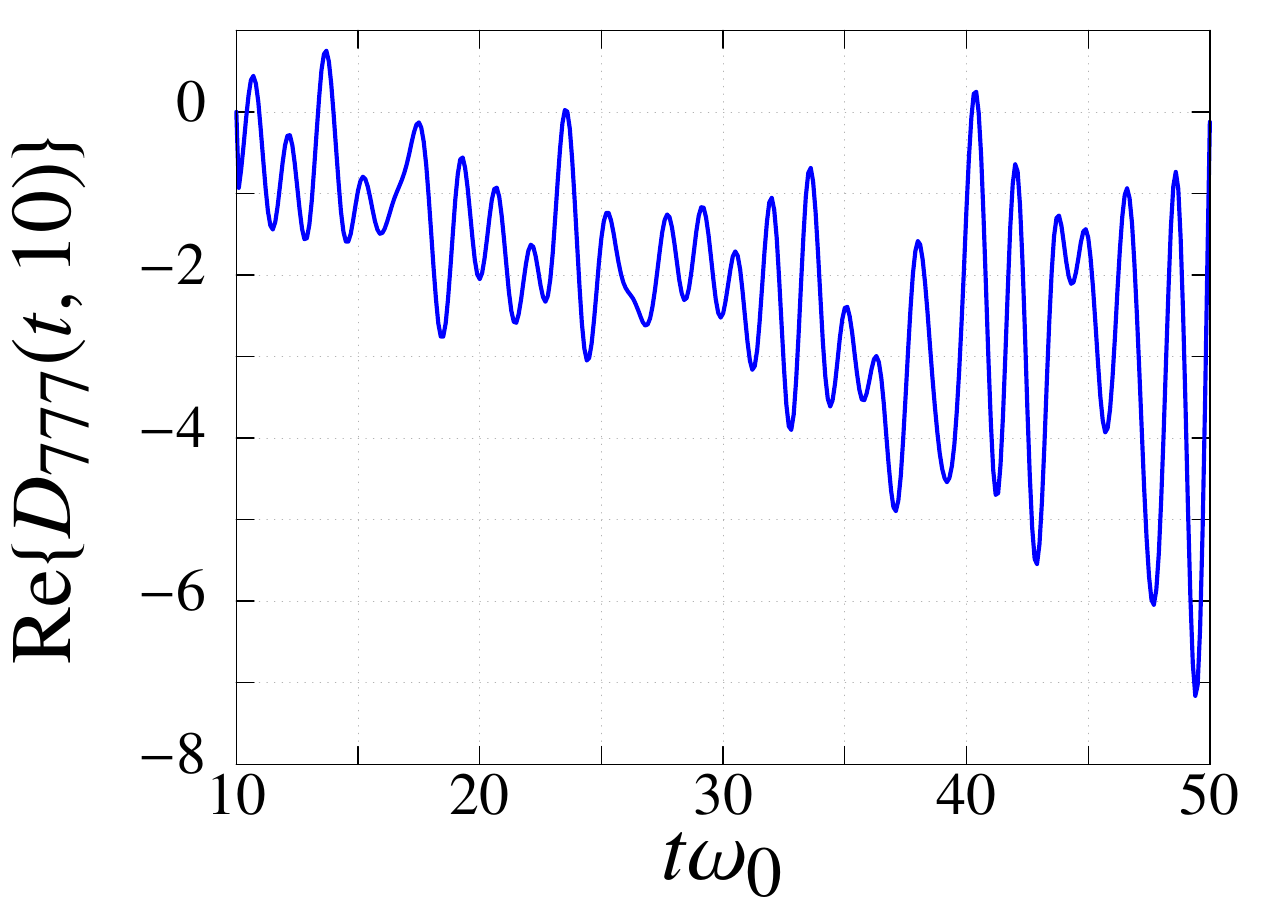}
\caption{
Real part of the three-body correlation function for the central spin in the 1D $XXZ$ spin lattice, $D_{777}(t, t')$, in arbitrary units at zero temperature ($\beta \hbar \omega_0 \to \infty$). The time $t'$ is fixed as $t' = 10$.
Note that the present three-body result is about $10^5$ times smaller than the two-body results in Fig.~\ref{fig:2corrTemp}.
\label{fig:3corr}
}
\end{figure}

First, we briefly examine the characteristics of the non-Gaussian noise that arises from the spin subenviroment. In the HSEOM approach, any correlation functions for the system can be evaluated by applying the system operators at each required time when conducting the time integration along the contour. \cite{NT18PRA} We then calculate the two-body and three-body correlation functions of the spins in the spin lattice coupled to the QT, $A_{jk}(t)=\langle [\hat \sigma_j^z (t), \hat \sigma_k^z(0)]\rangle$, $C_{jk}(t)=\langle \hat \sigma_j^z(t) \hat \sigma_k^z(0)+ \hat \sigma_k^z(0) \hat \sigma_j^z (t)) \rangle/2$, and  $D_{ijk}(t, t')=\langle \hat \sigma_i^z (t') \hat \sigma_j^z (t) \hat \sigma_k^z(0) \rangle$ for the fixed anisotropy $\Delta  = 1$ with the diagonal interaction Eq.~\eqref{eq:diag}. Here the expectation value is defined as $\braket{\hat O(t)} \equiv \sum_{\vec{n}} \braket{\psi_{\vec{n}} (t) | \hat{O}|\phi_{\vec{n}}(t)}$. Note that $D_{ijk}(t, t')$ is also optically observable, for example, in two-dimensional Raman spectroscopy.\cite{TM93JCP}

In Fig.~\ref{fig:2corrTemp}, we depict the time evolution of the antisymmetric correlation function of the central spin of the spin lattice $A_{77}(t)$ at zero and finite temperatures. For conventional Gaussian noise, this antisymmetric correlation function uniquely characterizes the dissipation and is independent of temperature. \cite{T06JPSJ,T14JCP,T15JCP,T20JCP} In the present case, however, we found that $A_{jk}(t)$ does depend on temperature.
Additionally, the noise is non-Gaussian, as can be seen from the three-body correlation function $D_{777}(t, t')$ illustrated in Fig.~\ref{fig:3corr}. 
This indicates that the antisymmetric correlation function is not the unique source that characterizes the dissipation, although the contribution of $D_{777}(t, t')$ is about $10^5$ times smaller than the two-body contribution in Fig.~\ref{fig:2corrTemp}. This is because the odd- and higher-order cumulant expansion terms of the noise correlation function contribute to the dynamics of the TLS.\cite{vankampen2007spp,toda2012statistical} 

\begin{figure}
\centering
\includegraphics[width=\linewidth]{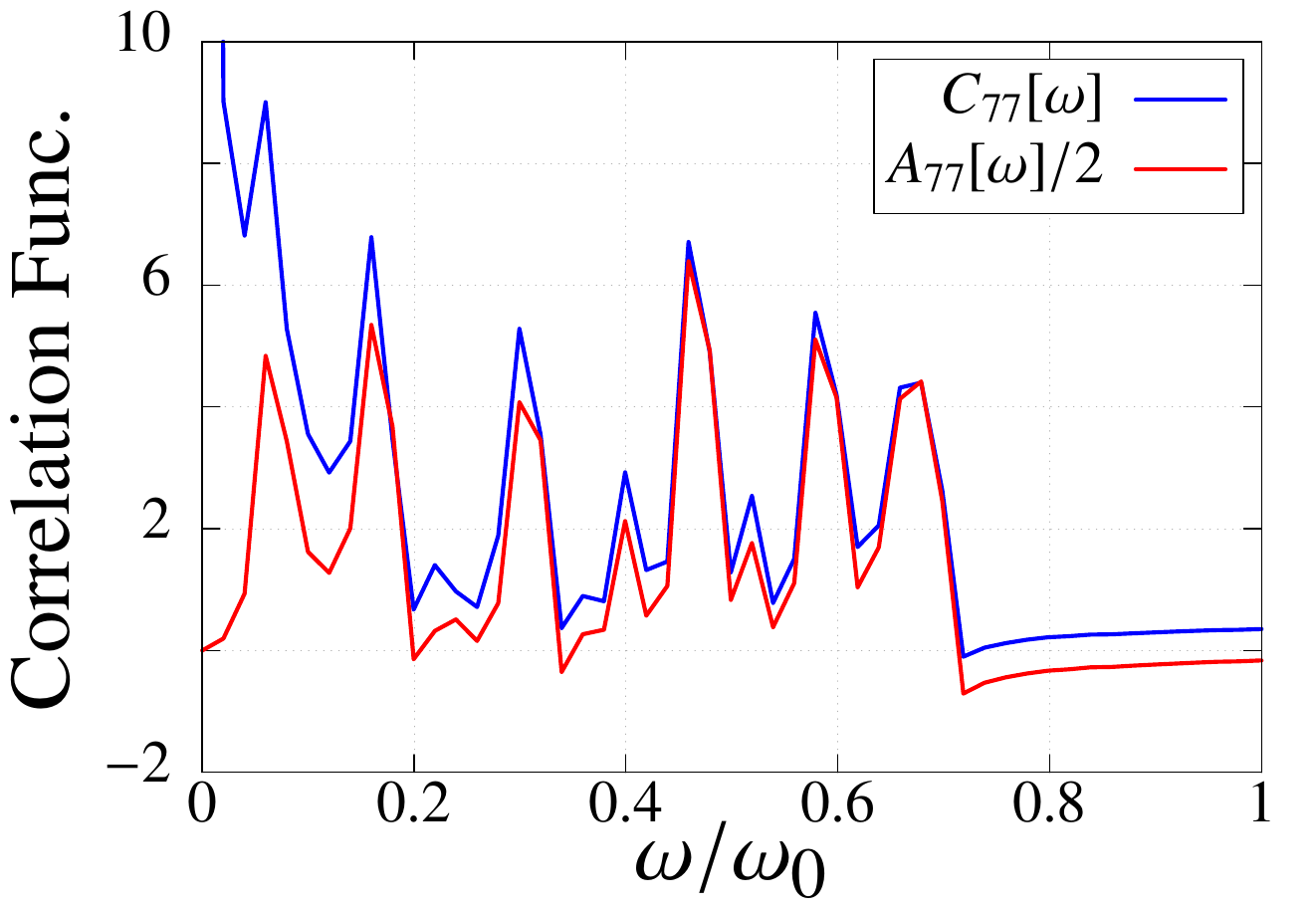}
\caption{
Fourier transforms of the symmetric and antisymmetric correlation functions, $C_{77}[\omega]$ and $A_{77}[\omega]/2$, in arbitrary units at zero temperature ($\beta \hbar \omega_0 \to \infty$).
\label{fig:2corrFDT}
}
\end{figure}

Next, we discuss the validity of the FDT for the present non-Gaussian case. Thus, we consider the Fourier transforms of the symmetric and antisymmetric correlation functions. Considering the time symmetry of $C_{jj}(t)$ and $A_{jj}(t)$, they can be defined as:
\begin{align}
C_{jj}[\omega] & = 2\int_{0}^{\infty} dt \cos \omega t \operatorname{Re}\{\braket{\hat{\sigma}_j^z(t)\hat{\sigma}_j^z(0)}\}, \\
A_{jj}[\omega] & = 4\int_{0}^{\infty} dt \sin \omega t \operatorname{Im}\{\braket{\hat{\sigma}_j^z(t)\hat{\sigma}_j^z(0)}\}.
\end{align}
Note that, as can be seen from Eqs.~\eqref{eq:CorrIm} and~\eqref{eq:CorrRe},  if the noise is Gaussian, the above functions satisfy the FDT expressed as:
\begin{align}
C_{jj}[\omega] = \frac{1}{2} \coth\left(\frac{\beta \hbar \omega}{2}\right) A_{jj}[\omega].
\end{align}
In Fig.~\ref{fig:2corrFDT}, we depict $C_{77}[\omega]$ and $A_{77}[\omega]/2$ at zero temperature ($\beta \hbar \omega_0 \to \infty$). To compute these correlation functions, we replace the upper bound of the Fourier integrals, $\infty$, with a finite time $T=50$. As a result, there is an artifact in the low-frequency region of the spectra, $\omega/\omega_0 \simeq 0$. So, we focus on the high-frequency region. There are various peaks corresponding to the different intersite spin--spin interactions. The peak intensities of  $C_{77}[\omega]$ and $A_{77}[\omega]/2$, however, do not agree at zero temperature ($\coth(\beta \hbar \omega/2) \to 1$) due to the non-Gaussian nature of the noise. This indicates the breakdown of the FDT. 
Because the noise perturbation from the TLS--spin interaction is weak in the present case, as the amplitude of the three-body correlation function indicated, and because the FDT is formulated within second-order perturbation theory, the discrepancy for the FDT is not significant.

\begin{figure}
\centering
\includegraphics[width=\linewidth]{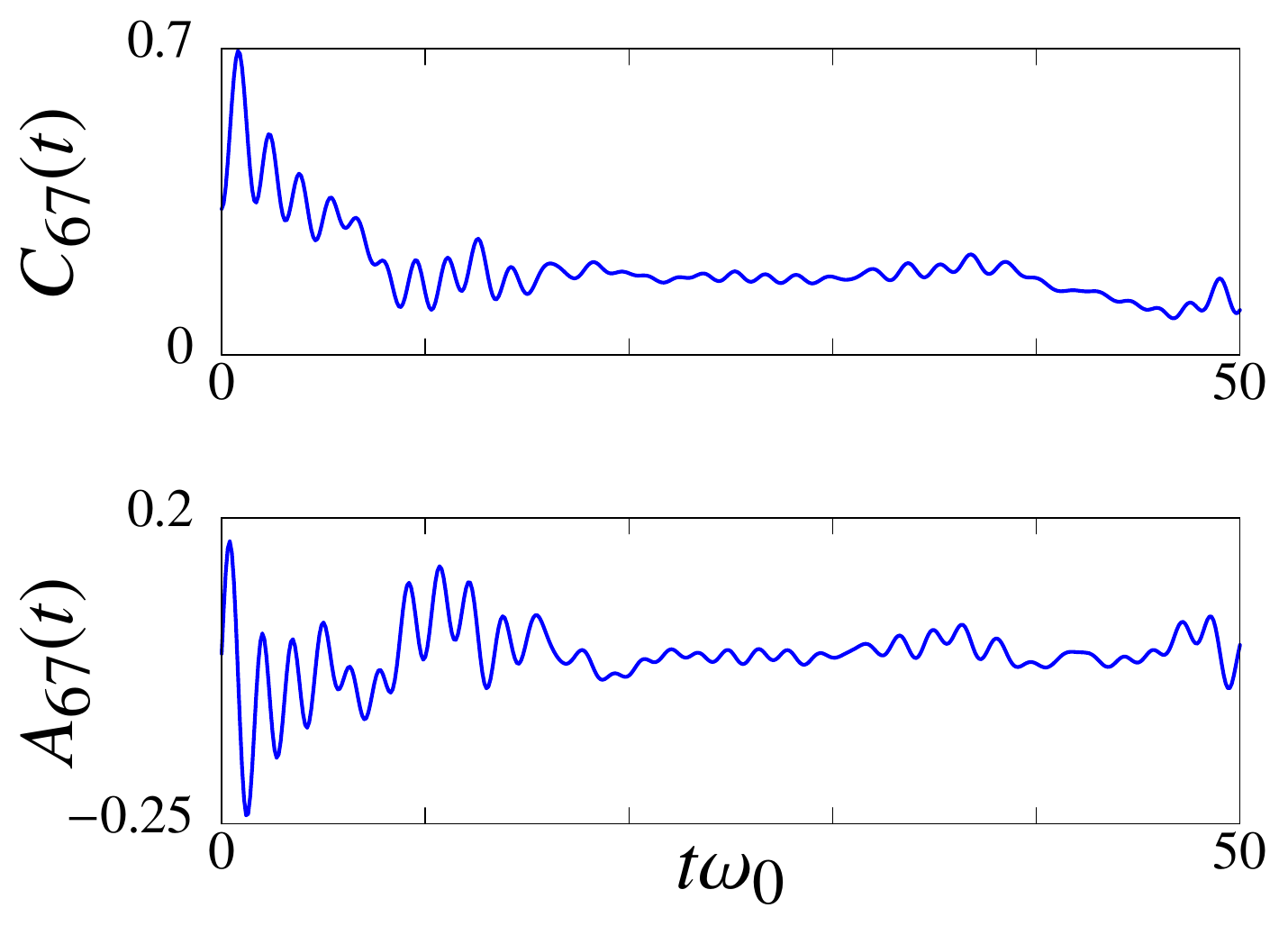}
\caption{
Symmetric and antisymmetric correlation functions for the intersite spins, $C_{67}(t)$ and $A_{67}(t)$, in arbitrary units at zero temperature ($\beta \hbar \omega_0 \to \infty$).
\label{fig:2corrIntersite}
}
\end{figure}

Finally, we examine the noise--noise correlation between different spin sites. We depict $A_{67}(t)$ and  $C_{67}(t)$ in Fig.~\ref{fig:2corrIntersite}. 
The noise at each site is strongly correlated because we assume that the single QT is coupled to all of the spins in the lattice and that the spins strongly interact with each other, as expressed in Eq.~\eqref{eq:xxz}.
The correlations are not easy to characterize, however, because they are controlled by complex spin--spin interactions that often undergo a phase transition at different temperatures. Because all the noise is generated in a quantum-mechanically consistent manner, the present model is useful for a rigorous investigation of quantum noise that arises from a complex environment. 

For details of the correlation functions with and without the QT, see Appendix~\ref{sec:appCorr}.

\subsubsection{Dissipative effects of the QT} \label{sec:thermo}

To characterize the dynamics of the TLS, we introduce the Loschmidt echo, defined as\cite{rossini07sb}:
\begin{align}
\mathcal{L}(t) = \frac{|\braket{+|\hat{\rho}_{S}(t)|-}|^2}{|\braket{+|\hat{\rho}_{S}(0)|-}|^2}.
\end{align}
Here, $\hat{\rho}_{S}(t)$ is the reduced density operator for the TLS:
\begin{align}
\hat{\rho}_{S}(t) & = \mathrm{tr}_{SE} \{\hat{\rho}_{\mathrm{S+SE}}(t) \} \nonumber \\
& = \sum_{\vec{n}}
\begin{bmatrix}
\braket{\psi^{u}_{\vec{n}}(t)|\phi^{u}_{\vec{n}}(t)} & \braket{\psi^{l}_{\vec{n}}(t)|\phi^{u}_{\vec{n}}(t)} \\
\braket{\psi^{l}_{\vec{n}}(t)|\phi^{u}_{\vec{n}}(t)} & \braket{\psi^{l}_{\vec{n}}(t)|\phi^{l}_{\vec{n}}(t)}
\end{bmatrix}
.
\label{eq:RDO}
\end{align}
The superscripts $u$ and $l$ in Eq.~\eqref{eq:RDO} represent the upper and lower halves of the AWFs
(we treat the operators in such a way that $\hat{O} = \hat{O}_{S} \otimes \hat{O}_{SE}$).

\begin{figure}[ht]
\centering
\includegraphics[width=\linewidth]{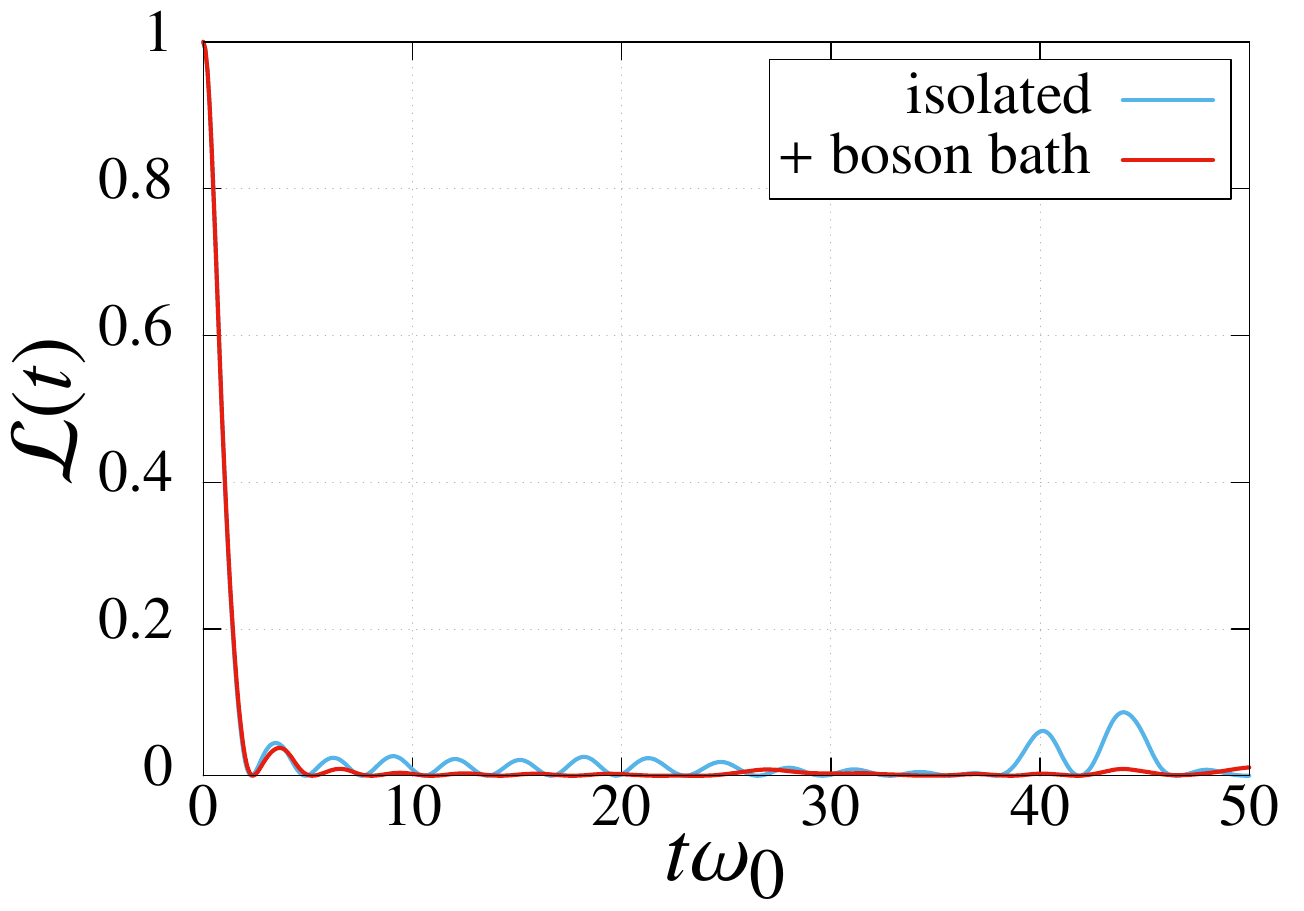}
\caption{Loschmidt echo $\mathcal{L}(t)$ of a TLS system that diagonally interacts with the subenvironment (spin lattice) in the absence (isolated, blue) and presence (+ boson bath, red) of the thermostat.
\label{fig:compare}
}
\end{figure}

First, we discuss the role of the QT when the TLS diagonally interacts with the spin-lattice subenviroment. Here, the population of the TLS does not change, so we do not present the results of the population relaxation. In Fig.~\ref{fig:compare}, we depict the Loschmidt echo with and without the QT. These two results agree in the short-time region. However, the echo peaks appear repeatedly in the case without QT (isolated case), while the recursive echo peaks decay quickly in the case with QT, due to the dissipation that arises from the~QT.

\subsubsection{Pure dephasing} \label{sec:pure}

\begin{figure}
\includegraphics[width=0.9\linewidth]{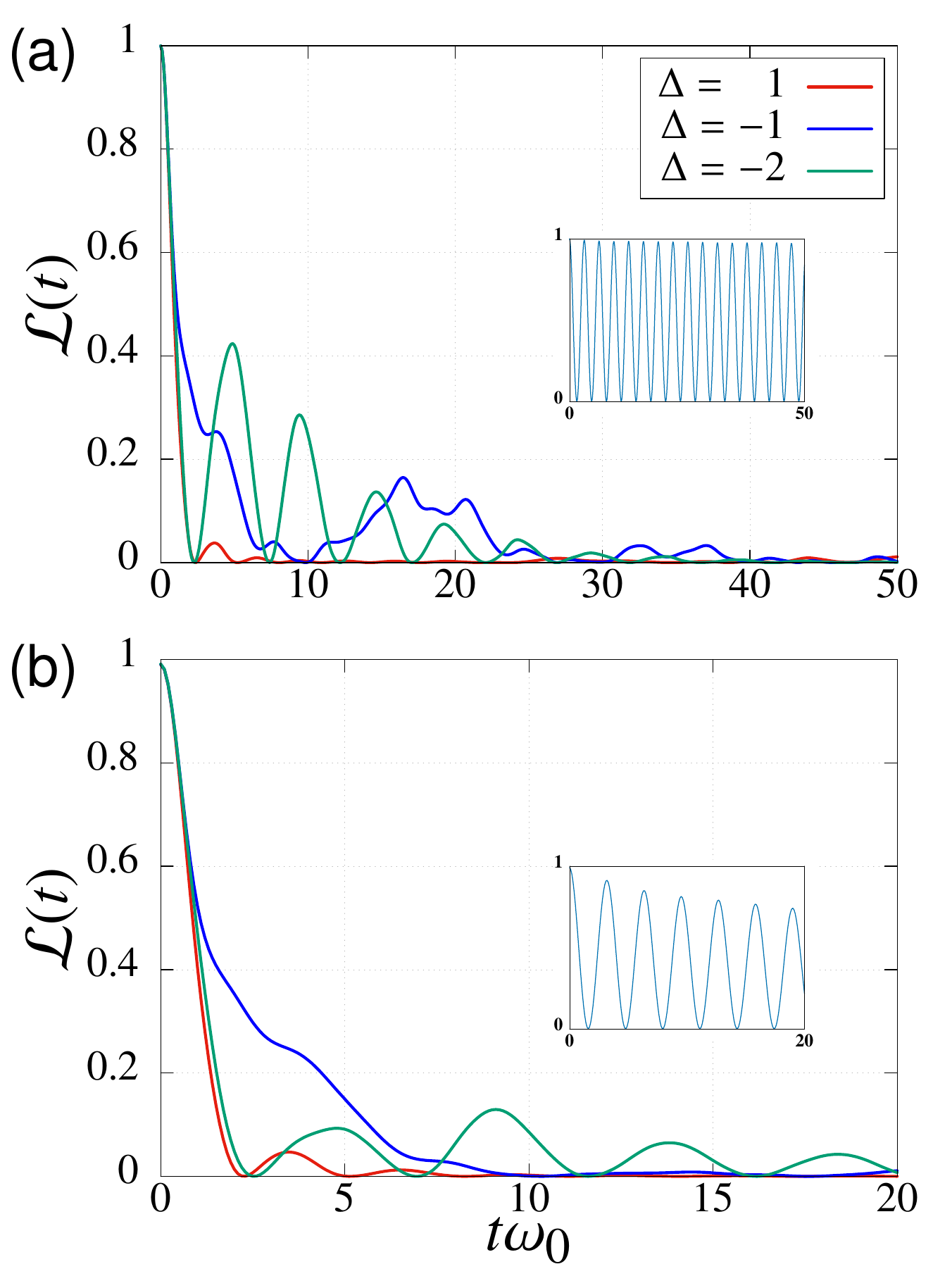}
\caption{
Loschmidt echo $\mathcal{L}(t)$ of a TLS that diagonally interacts with the spin lattice (pure dephasing) at (a) zero temperature ($\beta \hbar \omega_{0} \to \infty$) and (b) finite temperature ($\beta \hbar \omega_0 = 2$)  for $\Delta= 1$ (red), $-1$ (blue), and $-2$ (green). The inset displays the case for $\Delta = 2$.
\label{fig:pure10}
}
\end{figure}

In Figs.~\ref{fig:pure10}(a) and~\ref{fig:pure10}(b), we display the Loschmidt echo $\mathcal{L}(t)$ of a TLS that diagonally interacts with the spin lattice (pure dephasing) at (a) zero temperature and (b) finite temperature for various values of the spin anisotropy $\Delta$.
With weak TLS--spin interactions, the Loschmidt echo decays quickly when the $XXZ$ spin-lattice system is in the critical region ($|\Delta| \le 1$).\cite{haikka12z}  This is because the spins can be in different states without changing energy due to the highly degenerate spin states. Fast decoherence is then observed for $\Delta = \pm1$ at zero temperature [Fig.~\ref{fig:pure10}(a)], as reported previously for isolated systems.\cite{rossini07sb}   For the finite temperature in Fig.~\ref{fig:pure10}(b), the decoherence for $\Delta = \pm1$ is faster than that in Fig.~\ref{fig:pure10}(a), due to the relaxation that arises from the thermostat.

For isolated $XXZ$ systems, the Loschmidt echo never decays for $\Delta > 1$.\cite{rossini07sb}  This is also true for a zero-temperature QT, as illustrated in the inset of Fig.~\ref{fig:pure10}(a) for $\Delta =2$, because the ferromagnetic order of the spins is not altered by this weak QT interaction. At finite temperature, as illustrated in the inset of Fig.~\ref{fig:pure10}(b), the thermal fluctuation becomes large and thus, the echo decays gradually. 

An isolated spin-lattice system is in the antiferromagnetic phase for $\Delta < -1$. Although the echo signal does not decay in the perfectly antiferromagnetic regime for $\Delta \to -\infty$, it decays for finite $\Delta < -1$.\cite{rossini07sb} This holds even when the spin lattice is coupled to the thermostat, as illustrated in Figs.~\ref{fig:pure10}(a) and~\ref{fig:pure10}(b). 

Because the ground state becomes highly degenerate with the change $|\Delta| = 1 + \delta $
for $\delta \to 0$ (here $\delta$ is an infinitesimal), the profile changes significantly for $|\Delta| = 1$. This change at $\Delta = 1$ is similar to the first-order phase transition, while the change $\Delta = -1$ is moderate because the ground state energy and its derivatives with respect to $\Delta$ are continuous.\cite{franchini17}

\subsubsection{Population relaxation} \label{sec:relax}

\begin{figure}
\includegraphics[width=0.9\linewidth]{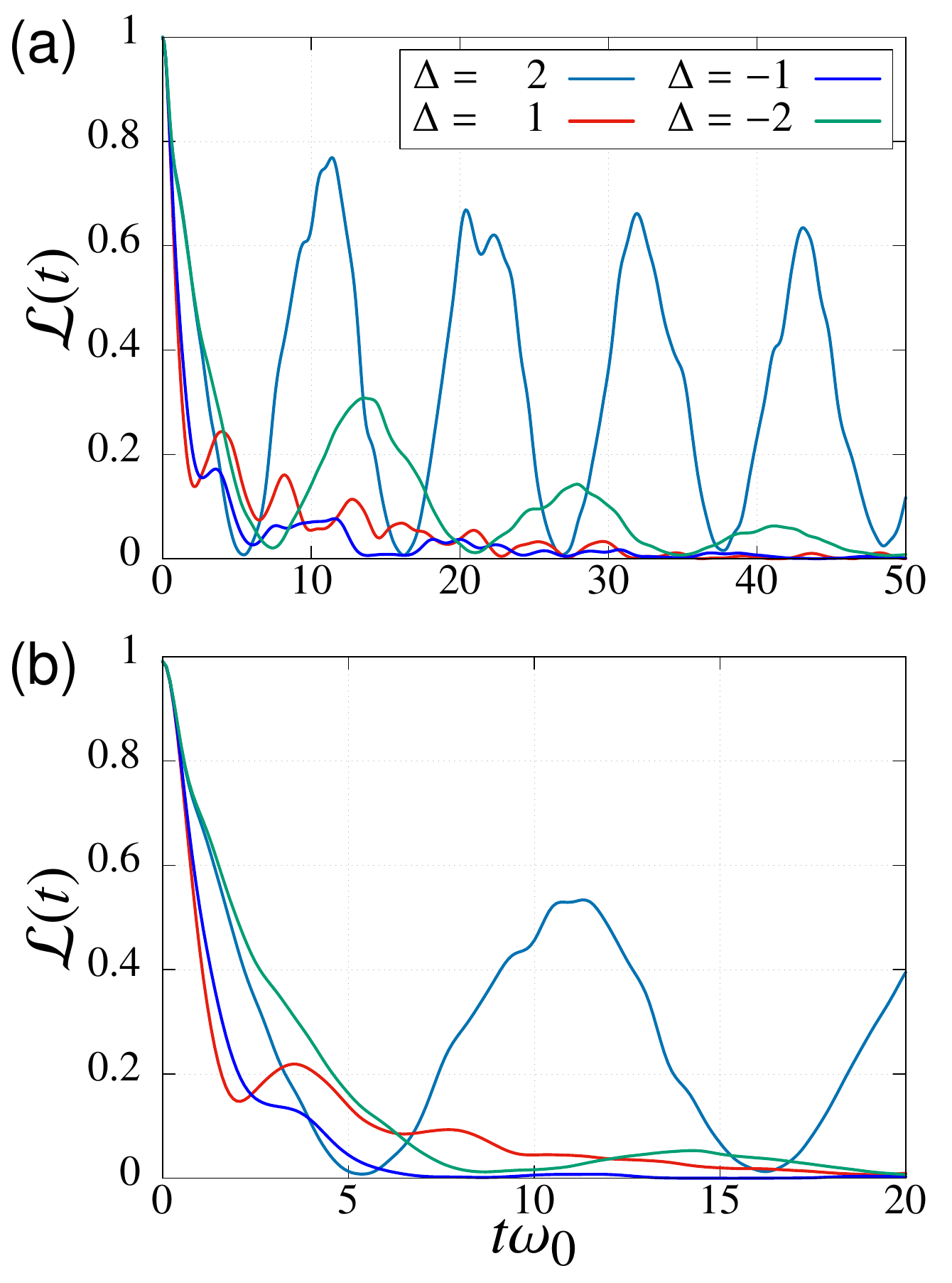}
\caption{
Loschmidt echo $\mathcal{L}(t)$ for a TLS that off-diagonally interacts with the spin lattice (population-relaxation case) at (a) zero temperature ($\beta \hbar \omega_{0} \to \infty$) and (b) finite temperature ($\beta \hbar \omega_0 = 2$).
\label{fig:relax10}
}
\end{figure}

\begin{figure}
\includegraphics[width=0.9\linewidth]{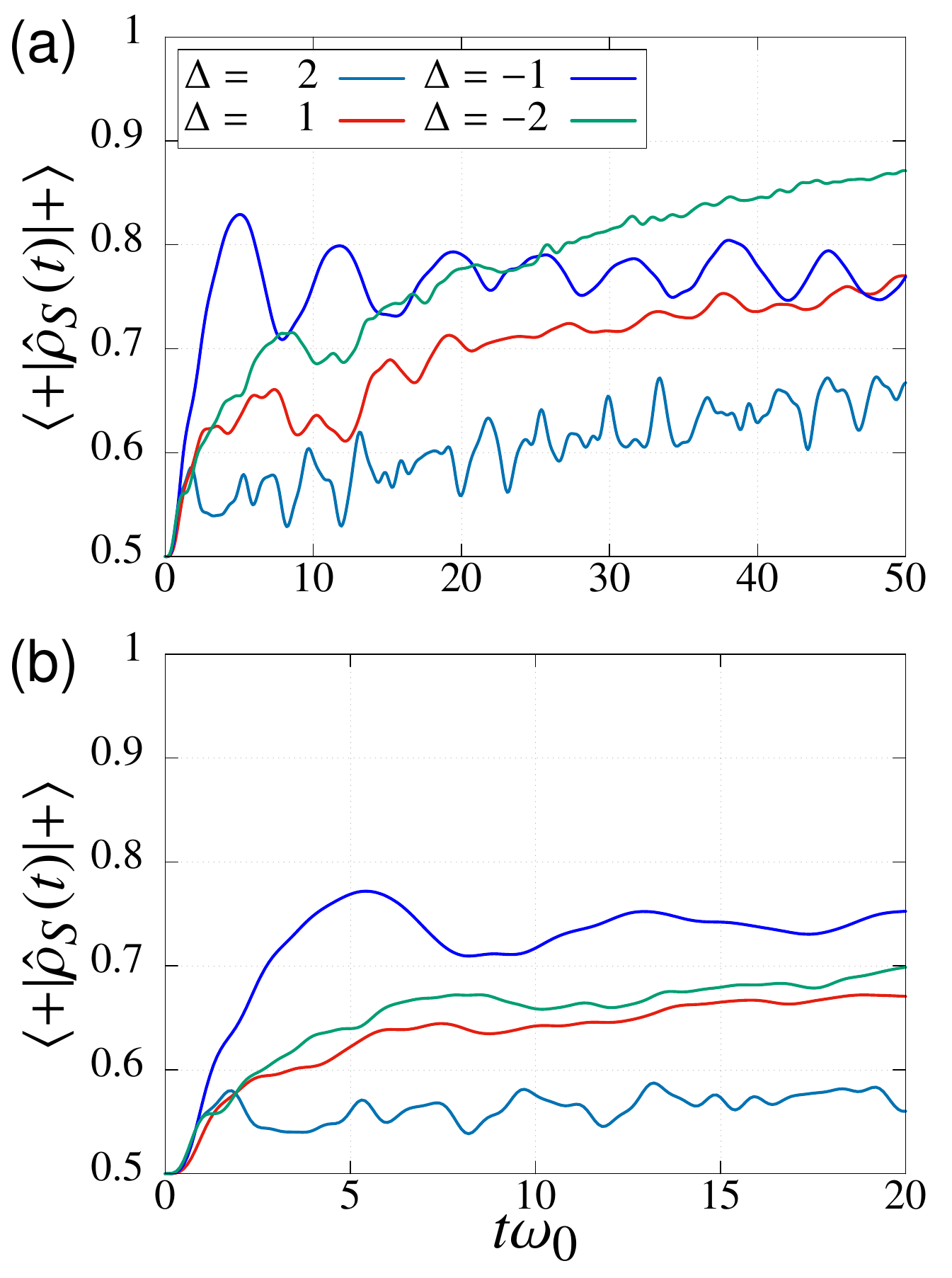}
\caption{
Ground-state population $\braket{+|\hat{\rho}_{S}|+}$ for a TLS that off-diagonally interacts with the spin lattice (population-relaxation case) at (a) zero temperature ($\beta \hbar \omega_0 \to \infty$) and (b) finite temperature ($\beta \hbar \omega_{0} = 2$).
\label{fig:relax00}
}
\end{figure}
  
Next, we display the Loschmidt echo $\mathcal{L}(t)$ and the ground-state population $\braket{+|\hat{\rho}_{S}(t)|+}$ for a TLS that off-diagonally interacts with the spin lattice (population-relaxation case) for various values of the spin anisotropy $\Delta$  at (a) zero and (b) finite temperatures in Figs.~\ref{fig:relax10} and~\ref{fig:relax00}, respectively.

In comparison with the pure dephasing (diagonal interaction) in Figs.~\ref{fig:pure10}(a) and~\ref{fig:pure10}(b), the recurrence interval of the echo peaks in Figs.~\ref{fig:relax10}(a) and~\ref{fig:relax10}(b) becomes longer in the population-relaxation case (off-diagonal interaction) when the $XXZ$ spin lattice is not in the critical region (i.e., $| \Delta| >1$).
This is due to the difference in coherence between the TLS and the subenvironment.
For pure dephasing, coherences such as $\braket{\hat{\sigma}_0^x \hat{\sigma}_{j_1}^z}$ are enhanced, whereas for the population-relaxation case, coherences such as $\braket{\hat{\sigma}_0^y \hat{\sigma}_{j_1}^x}$ and $\braket{\hat{\sigma}_0^y \hat{\sigma}_{j_1}^y}$ are enhanced.
Here, we focus on the difference in the direction of the spin at $j_1$.
The coherence of the TLS develops in the $z$ direction in the former case, whereas it develops in the $x$ and $y$ directions for the latter.  When $\Delta$ is large, the contribution from $\hat{\sigma}_j^z \hat{\sigma}_{j+1}^z$ becomes dominant in Eq.~\eqref{eq:xxz}.
Under this condition, the coherence between the TLS and spin lattice in the $z$ direction is not suppressed, while that in the $x$ and $y$ directions is suppressed due to the conversion of this coherence to higher-order correlations, for example, $\braket{\hat{\sigma}_0^y\hat{\sigma}_{j_1}^y\hat{\sigma}_{j_1+1}^z}$. This difference changes the recurrence interval time.

When $\Delta = \pm1$, the recurrence interval of the echo peaks changes slightly, because all of the spin--spin interaction terms in Eq.~\eqref{eq:xxz} contribute equally, which leads to delocalization of the coherences in all directions in the spin lattice.

In Figs.~\ref{fig:relax00}(a) and~\ref{fig:relax00}(b), we depict the time evolution of the population. Due to the energy relaxation from the TLS to the subenviroment, the ground-state population grows as a function of time.

Apart from the contribution of the equilibrium population, $\braket{+|\hat{\rho}_{S}(t \to \infty)|+}$, the population relaxation is faster in the critical case ($\Delta = \pm 1$) than when $\Delta = \pm 2$. In both the pure-dephasing and population-relaxation cases, the relaxation process was enhanced in the critical region. A prominent difference in the population dynamics is observed between the zero and finite temperatures when $\Delta = -2$, as the spin lattice is in the antiferromagnetic phase.

\subsubsection{Size effects of the subenviroment} \label{sec:size}

\begin{figure}
\centering
\includegraphics[width=0.9\linewidth]{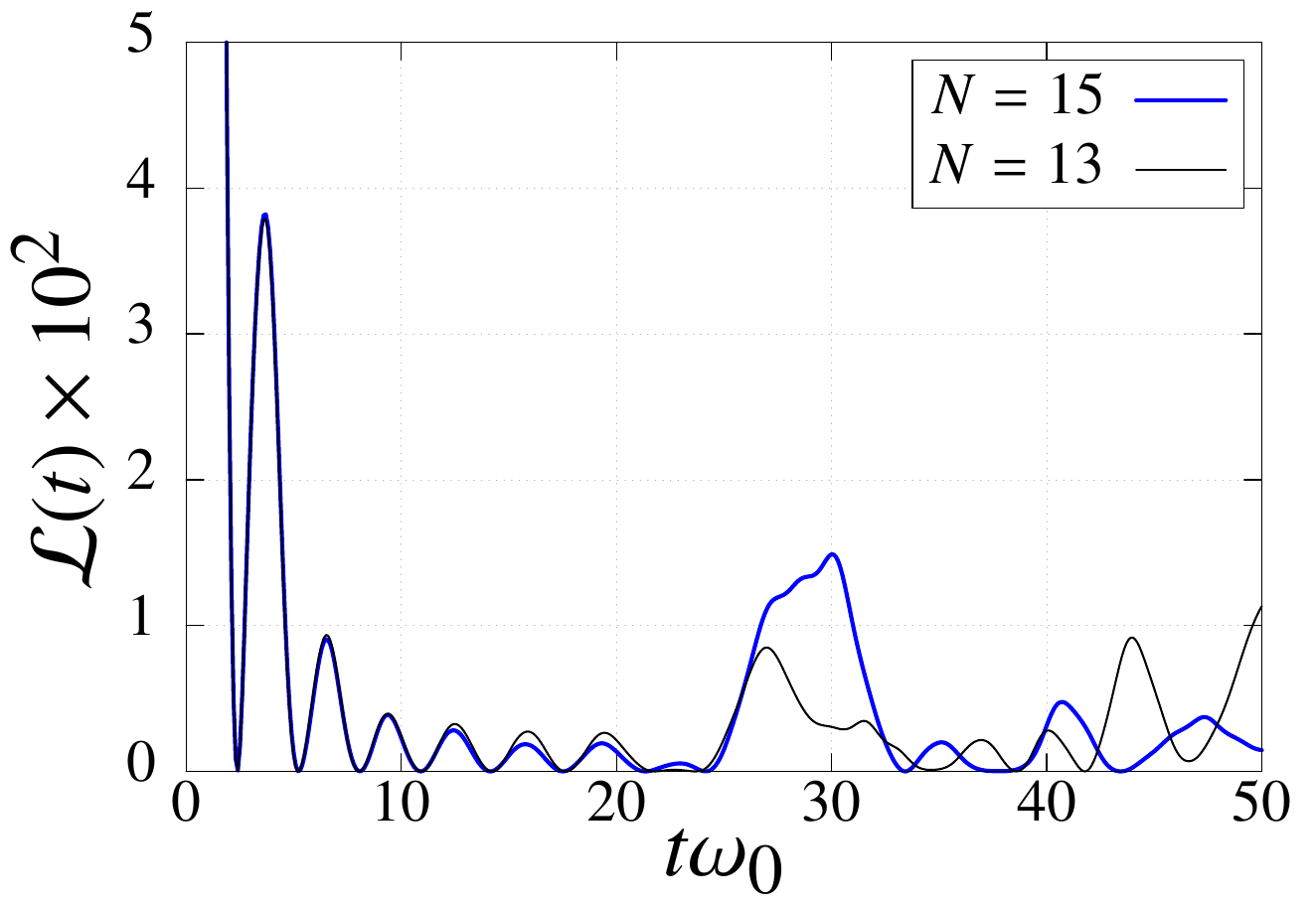}
\caption{
Loschmidt echo $\mathcal{L}(t)$ of a TLS that diagonally interacts with the spin lattice (pure dephasing) consisting of $15$ spins. The inverse temperature and spin anisotropy are set as $\beta \hbar \omega_{0} \to \infty$ and $\Delta = 1$, respectively. The Loschmidt echo for $\Delta=1$ in the 13-spin case in Fig.~\ref{fig:pure10}(a) is replotted for reference.
\label{fig:pure16}
}
\end{figure}

\begin{figure}
\centering
\includegraphics[width=0.85\linewidth]{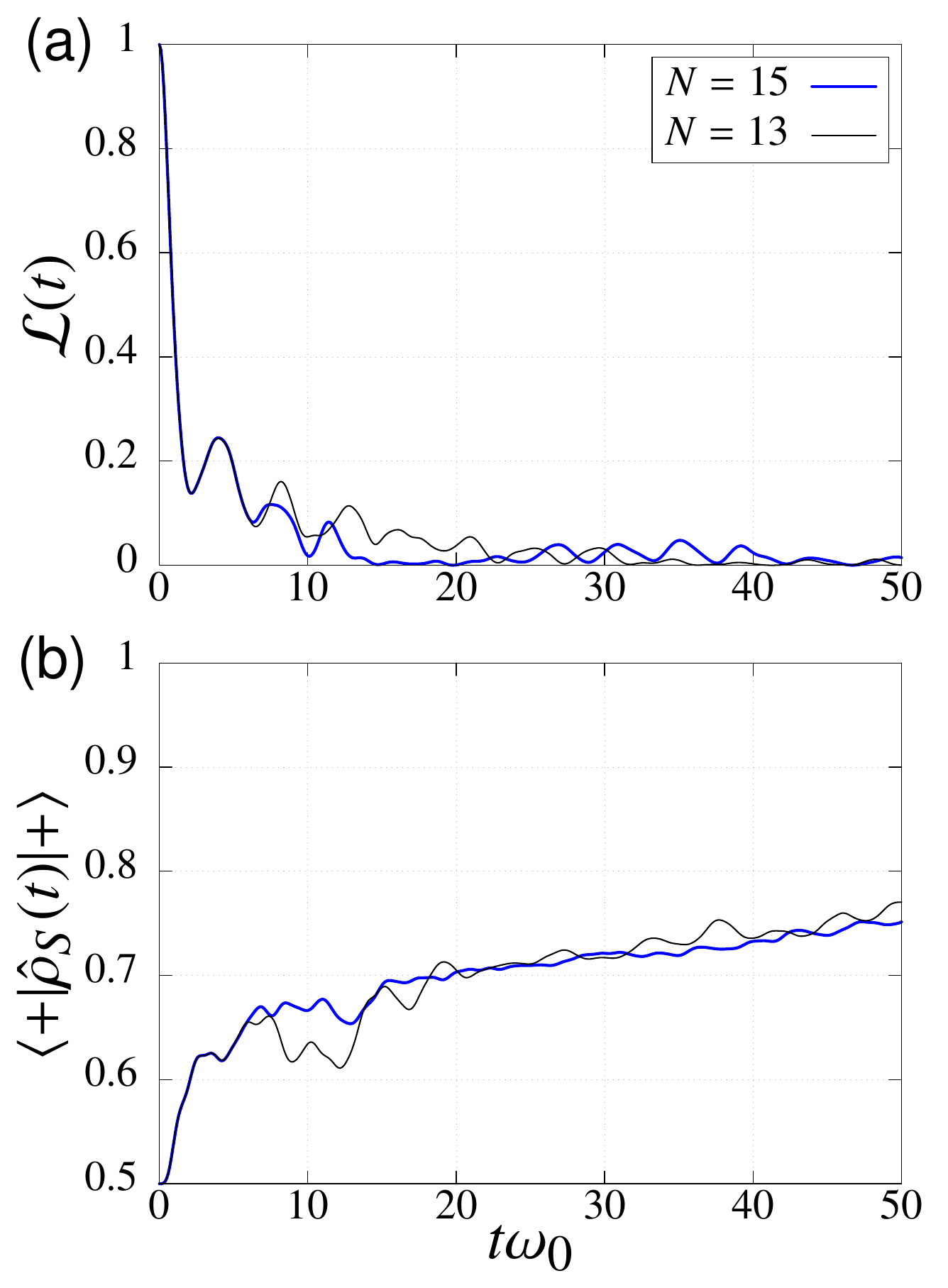}
\caption{
(a) Loschmidt echo $\mathcal{L}(t)$ and (b) ground-state population $\braket{+|\hat{\rho}_{S}|+}$ of a TLS that off-diagonally interacts with the spin lattice (population-relaxation case) consisting of $15$ spins.
The temperature and anisotropy are set as zero ($\beta \hbar \omega_{0} \to \infty$) and $\Delta = 1$, respectively. The Loschmidt echo and the ground-state population for $\Delta=1$ in the 13-spin case in Figs.~\ref{fig:relax10}(a) and~\ref{fig:relax00}(a) are replotted for reference.
\label{fig:relax16}
}
\end{figure}

Finally, we discuss the size effects of the spin lattice. Figures~\ref{fig:pure16} and~\ref{fig:relax16} display the time evolution of the Loschmidt echo and the population in the pure-dephasing and population-relaxation cases for 13-spin (black curves) and 15-spin (blue curves) lattices. In these figures, the 13-spin results for $\Delta=1$ in Figs.~\ref{fig:pure10}(a),~\ref{fig:relax10}(a), and~\ref{fig:relax00}(a) are replotted for reference.
In the pure-dephasing case in Fig.~\ref{fig:pure16}, the calculated results for the $13$-spin and $15$-spin Loschmidt echoes are qualitatively the same up to time $t \omega_0=20$, after which a difference arises in that the echo in the $15$-spin case exhibits slower recursive motion  ($T\omega_0 \simeq 30$) than in the $13$-spin case ($T\omega_0 \simeq 25$) due to the size effects of the quantum revival.

In the population-relaxation case in Fig.~\ref{fig:relax16}, 
the time evolution of the echo signals matches only up to $t\omega_0 \simeq 7$. 
Furthermore, the Loschmidt echo decays faster in the $15$-spin case than in the $13$-spin case.
These differences are due to the different mechanisms of decoherence and population relaxation, as known from photon echo measurements in ultrafast nonlinear spectroscopy.

While we observe good agreement of the Loschmidt echo signal up to time $t\omega_0=20$ in the $13$- and $15$-spin cases, we found that the rephasing echo peaks around $t \omega_0=25$--30. This is an artifact of the lattice size. This peak can be suppressed by either increasing the lattice size or increasing the QT coupling strength.  
In the population-relaxation case, the energy relaxation, as well as the dephasing, occurs during the time evolution, and the effects of the thermostat are much larger than for pure dephasing. Therefore, the Loschmidt echo in Fig.~\ref{fig:relax16}(a) decays as the population of the excited state decreases, as can be seen from Fig.~\ref{fig:relax16}(b). The signal in the 15-spin case decays faster than in the 13-spin case, because, in the present model, all the spins in the lattice are coupled to the thermostat and the effective coupling strength becomes stronger as the spin system becomes larger.
Although the time evolution of the population is qualitatively the same in the longtime regime, the amplitude of the oscillations in the short-time regime ($t \omega_0 \simeq 10$) is larger in the $13$-spin case.

In realistic situations, it is impossible for a spin lattice or any other subenvironment, whether small or large, to exist in isolation. Thus, it is essential to consider some kind of heat source for the subenvironment, as we do in the present paper. Because the QT indirectly and intricately influences the subenvironment, sufficient care must be taken in modeling from an environment with infinite degrees of freedom to a subenvironment with~a~QT.

\section{CONCLUDING REMARKS} \label{sec:conclude}

In this paper, we introduced a subenvironment system coupled to a QT to describe the dynamics of a system in a complex non-Gaussian environment in a quantum-mechanically consistent manner. To accurately simulate the dynamic and thermal aspects of the complex environment in the framework of the present approach, the coupling strength between the subenvironment and QT must be weak and there must be many subenvironmental degrees of freedom. We employed the wavefunction-based HSEOM method to handle the large number of subenvironmental degrees of freedom and the effects of the QT in a numerically rigorous manner. While the thermal noise that arises from the QT is Gaussian, the noise generated from the subenvironment is non-Gaussian. Thanks to the HSEOM formalism, we can accurately treat the quantum entanglement between the subenvironment and the QT. Moreover, because we are simulating the subenvironmental degrees of freedom explicitly, we can treat the quantum entanglement between the main system and subenvironment + QT precisely.

To demonstrate our approach, we simulated the time-irreversible dynamics of a TLS that interacts with a 1D $XXZ$ spin-lattice system coupled to a QT. We found that the decoherence is enhanced when the spin subenvironment is in the critical region, even at finite temperature, for both the pure-dephasing and population-relaxation cases. The significant difference between the isolated system and the present system is that the Loschmidt echo decays at finite temperature, even when the spin lattice is out of the critical region. 
We showed that the noise generated from the spin lattice is non-Gaussian and non-local. Moreover, this noise does not satisfy the FDT, while the steady-state solution with this noise is the thermal equilibrium state of the total system, as the HSEOM approach guarantees.
Further work to find the relation between fluctuations and dissipation of the non-Gaussian and non-local noise should be conducted by evaluating the higher-order cumulant of the noise correlation function.  Along this direction, it should also be interesting to examine the effects of fermionic Gaussian noise by replacing $\coth(\beta \hbar \omega/2)$ in Eq.~\eqref{eq:CorrRe} with $\tanh(\beta \hbar \omega/2)$. The effects of multiple QTs, which may suppress the quantum coherence of the subenviroment, should also be examined.\cite{NT21JCP}

Introducing a QT may be regarded as a realistic approach to describing the inherent dissipative features of a spin lattice or any other subenvironment. We found, however, that the quantum coherence of the subenviroment system is spatially correlated in a relatively wide region, and thus, we need to consider the largest number of degrees of freedom that we can handle numerically. 
For a simulation of a $20$-spin system using our computer program for the HSEOM, we need about 130 gigabytes of RAM. Such computations should be facilitated by advances in formulations and hardware such as a graphics processing unit and massively parallel computers. Extending the present method to larger systems will be a future challenge.

\section*{ACKNOWLEDGMENTS}
Y.T. was supported by a Kakenhi grant from the Japan Society for the Promotion of Science  (Grant No.~B 21H01884).

\appendix

\section{Derivation of HSEOM} \label{sec:HSEOM}

Instead of treating the thermal noise as a random variable, here we introduce a harmonic heat bath as a noise source in a spin system described by Eq.~\eqref{eq:SpinH}. The total Hamiltonian is\cite{caldeira83cl,CLModel, LeggettRMP87A}
\begin{align}
\hat{H}_{\mathrm{tot}} = \hat{H}_{\mathrm{S+SE}} + \hat{H}_{\mathrm{QT}}.
\label{eq:totH}
\end{align}
Here, we assume that all of the spins in the spin lattice are coupled to the single harmonic bath.

The matrix elements $\braket{n|\hat{\rho}_{\mathrm{S+SE}}|m}$ are evaluated with a path integral along the contour in the following form:
\begin{align}
\braket{n|\hat{\rho}_{\mathrm{S+SE}}|m} = & \int \frac{d z'_{i}}{\mathcal{N}} \int \frac{d z_{i}}{\mathcal{N}} 
\int \frac{d z'_{f}}{\mathcal{N}} \int \frac{d z_{f}}{\mathcal{N}} \nonumber \\ 
& \times \int _{C = z_{i} \to z_{f} \to z'_{f} \to z'_{i}} \hspace{-6em}\mathcal{D} [\tilde{z}(\cdot)]
\phi_{n}^{*}(z_{f}, t) \phi_{m}(z'_{f}, t)  \nonumber \\
& \times \exp\left[\frac{i}{\hbar} \int_{C} d\tau L_{\mathrm{S+SE}}(\dot{\tilde{z}}, \tilde{z};\tau)\right] \mathcal{F}(\tilde{z}) \nonumber \\
& \times \braket{z_{i}|\hat{\rho}_{\mathrm{S+SE}}(0)|z'_{i}},
\end{align}
where $L_{\mathrm{S+SE}}(\dot{\tilde{z}}, \tilde{z};\tau)$ is the Lagrangian for the spin system in terms of the boson-coherent, fermion-coherent, and spin-coherent, and displacement representations (the normalization factor $\mathcal{N}$ depends on the representation of $z$).
The number of degrees of freedom for the bosonic heat bath are reduced to the influence functional along the contour, $\mathcal{F}$, whose form is expressed below.
The contour (path) integral, $\int_{C} d\tau$ ($\int_{C} \mathcal{D} [\tilde{z}(\cdot)]$), is realized with the aid of the projection operator, $\phi_{n}^{*}(z_{f}, t) \phi_{m}(z'_{f}, t) = \braket{n|z_{f}}\braket{z'_{f}|m}$, and $z$ along the contour is represented by $\tilde{z}$.

The influence functional can be obtained analytically because of the harmonicity of the heat bath.
By means of the two-time correlation function of the bosonic heat bath, $\alpha(t) = \hbar \int_{0}^{\infty} d \omega \mathcal{J}(\omega) [\coth(\beta \hbar \omega/2) \cos \omega t - i \sin \omega t]$, the influence functional can be expressed as follows:\cite{T14JCP,T15JCP} 
\begin{align}
\mathcal{F}(\tilde{z}) = \exp\left[-\frac{1}{\hbar^2}\int_{C} d\tau \int_{C'} d\tau' V(\tilde{z};\tau) \alpha(\tau - \tau') V(\tilde{z};\tau)\right].
\end{align}
Here $C'$, along which the integration over $\tau'$ is carried out, is the contour $C$ up to $\tau$.

If the Ohmic spectral density has a circular cutoff, as in Eq.~\eqref{eq:circular}, then the imaginary part of the bath correlation function defined by $\alpha(t)=\alpha'(t)-i\alpha''(t)$ can be analytically evaluated:\cite{Ando98}
\begin{align}
\alpha''(t) =  c_1 J_{1}(\nu t) + c_3 J_{3}(\nu t),
\end{align}
where $c_1=c_3=\pi \hbar \zeta \nu^{2}/{8}$. In the high-temperature limit, $\beta \to 0$, the real part of the bath correlation function reduces to $\alpha'(t) = {\pi} {\zeta \nu}(J_{0}(\nu t) + J_{2}(\nu t))/2\beta$.
This indicates that the circular cutoff is suitable for constructing the HSEOM.

By using special (and orthogonal) functions characterized by the differential equations $d \varphi_{k}(t) / dt = \sum_{k'} \eta_{k, k'} \varphi_{k'}(t) $ for the two-time correlation function of the bosonic heat bath, then since $\alpha(t) = \sum_{k} c_{k} \varphi_{k}(t)$, we can derive the HSEOM in the following form:\cite{NT18PRA} 
\begin{align}
\frac{\partial}{\partial s}  \ket{\Phi_{\vec{n}} (s; n'_{i})} = &
\mp \frac{i}{\hbar} \hat{H}_{\mathrm{S+SE}} \ket{\Phi_{\vec{n}} (s; n'_{i}) } \nonumber \\
&
\pm \sum_{k=0}^{K-1} \sum_{k'=0}^{K-1} \eta_{k, k'} n_{k}
\ket{ \Phi_{\vec{n} - \vec{e}_{k} + \vec{e}_{k'}}  (s; n'_{i})} \nonumber \\
& \mp \frac{i}{\hbar} \hat{V} \sum_{k=0}^{K-1} c_{k} \ket{  \Phi_{\vec{n} + \vec{e}_{k}}  (s; n'_{i})} \nonumber \\
&
\mp \frac{i}{\hbar} \hat{V} \sum_{k=0}^{K-1} n_{k} \varphi_{k} (0)
\ket{  \Phi_{\vec{n} - \vec{e}_{k}}  (s; n'_{i})}.
\label{eq:HSEOM}
\end{align}
Here, we restrict the number of special functions to $K$ and approximate the differential equations and two-time correlation function.
The vector $\vec{n} = [n_{0}, \ldots, n_{K-1}]$, which consists of non-negative integers, distinguishes the AWFs. It is in the following form:
\begin{gather}
\ket{ \Phi_{\vec{n}} (s; n'_{i}) } =
\sum_{n, n_{i}} \ket{n} \int \frac{dz}{\mathcal{N}} \braket{n|z}   \int \frac{dz_{i}}{\mathcal{N}} 
\int _{\tilde{z}(\tau(0))}^{\tilde{z}(\tau(s))} \hspace{-2em}\mathcal{D} [\tilde{z}(\tau(\cdot))] \nonumber \\
\begin{aligned}[b]
& \times \{\theta(t - s) + \theta(s - t) \phi_{n}^{*}(z_f, t) \phi_{m}(z'_f, t) \}\\
&\times \prod_{k=0}^{K-1}
\left(\hspace{-1pt}-\frac{i}{\hbar} \int _{0}^{s} \hspace{-7.5pt}ds'' \frac{d\tau(s'')}{d s''}
\varphi_{k}(\tau(s)-\tau(s'')) V(\tilde{z}, \tau(s'')) \hspace{-1pt} \right) ^{n_{k}} \\
&\times \exp\left[ \frac{i}{\hbar} \int _{0}^{s} ds' \frac{d\tau(s')}{ds'}
L_{\mathrm{S+SE}}(\dot{\tilde{z}}, \tilde{z}, \tau(s'))  \right] \mathcal{F}(s, V) \\
&\times
\braket{z_{i}|n_{i} } \braket{ n_{i} |\hat{\rho}_{S}(0)| n'_{i} }.
\label{eq:AWF}
\end{aligned}
\end{gather}
The vector $\vec{e}_{k}$ is the unit vector in the $k$th direction.
In Eq.~\eqref{eq:AWF}, $\theta(t)$ is the step function, and the influence functional along the contour $\mathcal{F}(s, V)$ is defined as follows:
\begin{gather}
\mathcal{F} (s, V) = \exp \left[-\frac{1}{{\hbar}^2} \int_{0}^{s} ds' \frac{d\tau(s')}{ds'}
\int_{0}^{s'} ds'' \frac{d\tau(s'')}{ds''} \right. \nonumber \\
\left. \times V(\tilde{z}, \tau(s')) \sum_{k=0}^{K-1} c_{k} \varphi_{k}(\tau(s')-\tau(s'')) V(\tilde{z}, \tau(s''))  \right].
\end{gather}
We have introduced the time variable $\tau(s)$ for $0 \leq s \leq 2t$, defined as:
\begin{equation}
\tau(s) \equiv
\begin{cases}
  s, & 0 \leq s \leq t\\
  2t-s, & t \leq s \leq 2t.
\end{cases}
\end{equation}

The HSEOM expressed in Eq.~\eqref{eq:HSEOM} have a disadvantage that each line integral is carried out independently, and calculations for the time evolution of reduced density matrices are time-consuming.
We remove this limitation by applying the method developed by Xu \textit{et al.}\cite{xu11heomheis}
We utilize that the matrix elements $\braket{n|\hat{\rho}_{\mathrm{S+SE}}|m}$ are described by the following equation after appropriate truncation of the AWFs:
\begin{align}
\braket{n|\hat{\rho}_{\mathrm{S+SE}}|m} = \langle \braket{\Phi|e^{\bm{\Lambda}_{l}t}|m}\braket{n|e^{\bm{\Lambda}_{u}t}|\Phi} \rangle,
\end{align}
where the operators $\bm{\Lambda}_{u}$ and $\bm{\Lambda}_{l}$ correspond to the upper and lower signs on the right-hand side of Eq.~\eqref{eq:HSEOM}, respectively.
The vector $\ket{\Phi}\rangle = [\ket{\Phi_{\vec{0}}(s=0)} = \ket{\psi}, \ket{\Phi_{\vec{e}_{0}}(s=0)} = 0, \ldots]$ is the vector in which the AWFs are aligned.
Here, we consider the initial states, $\hat{\rho}_{\mathrm{S+SE}}(0) = \ket{\psi}\bra{\psi}$.
In terms of the vectors, $\ket{\Phi(t)}\rangle= e^{\bm{\Lambda}_{u}t} \ket{\Phi}\rangle $ and $\ket{\Psi(t)}\rangle  = e^{\bm{\Lambda}^{\dagger}_{l} t} \ket{\Phi} \rangle$, we derive the modified HSEOM, in which the computation time is linear with $t$, as Eqs.~\eqref{eq:SEOMnewA} and~\eqref{eq:SEOMnewB}.

\section{Correlation functions with and without a QT} \label{sec:appCorr}

\begin{figure}
\centering
\includegraphics[width=\linewidth]{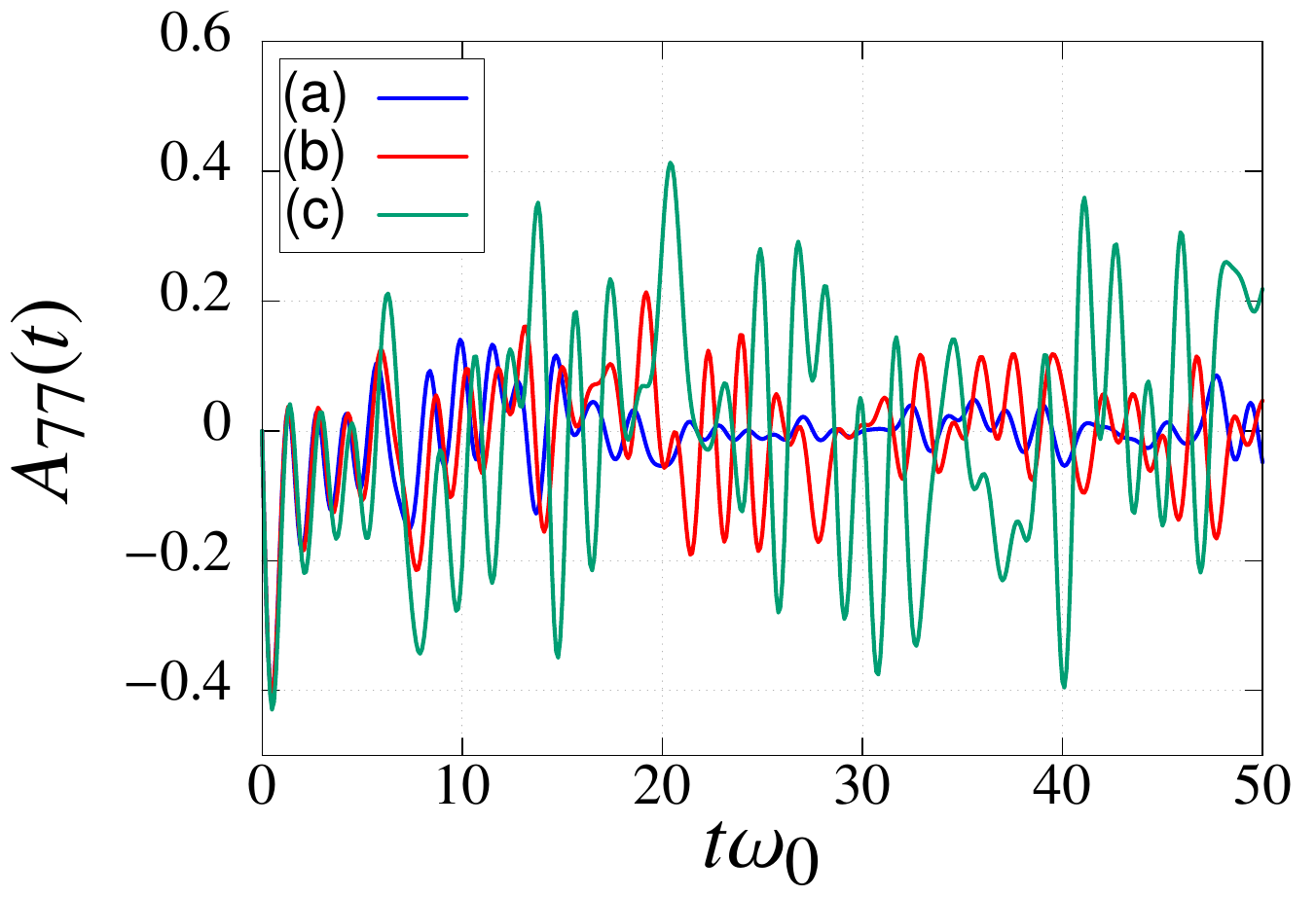}
\caption{
Antisymmetric correlation function $A_{77}(t)$ for the central spin in a 1D $XXZ$ spin lattice in arbitrary units. The blue, red, and green curves are for (a) a TLS + spin lattice + QT, (b) a TLS + spin lattice, and (c) an isolated spin lattice. The curve (a) is replotted from Fig.~\ref{fig:2corrTemp} for $\beta \hbar \omega_0 \to \infty$.
We cannot define the temperature in cases (b) and (c) because those systems do not couple with the QT.
\label{fig:2corrWO}
}
\end{figure}

In this appendix, we discuss the difference in the correlation functions for the lattice spins with and without the QT. The anisotropy is fixed as $\Delta = 1$.
In Fig.~\ref{fig:2corrWO}, the antisymmetric correlation functions $A_{77}(t)$ for (a) a TLS + spin lattice + QT  ($\epsilon/\omega_0 = 1$, $\hbar \zeta = 0.01$, $\beta \hbar \omega_0 \to \infty$, blue), (b) a TLS + spin lattice  ($\epsilon/\omega_0 = 1$, $\hbar \zeta = 0$, red), and (c) an isolated spin lattice ($\epsilon/\omega_0 = 0$, $\hbar \zeta = 0$, green) are depicted.
We found that the interactions with the TLS as well as with the QT suppress the oscillatory amplitude of the correlation function.
The frequencies of oscillation in cases (a), (b), and (c) are almost the same, indicating that the characteristic feature of the two-body correlation function of the central spin is predominantly determined by the spin-lattice system.

\begin{figure}
\centering
\includegraphics[width=\linewidth]{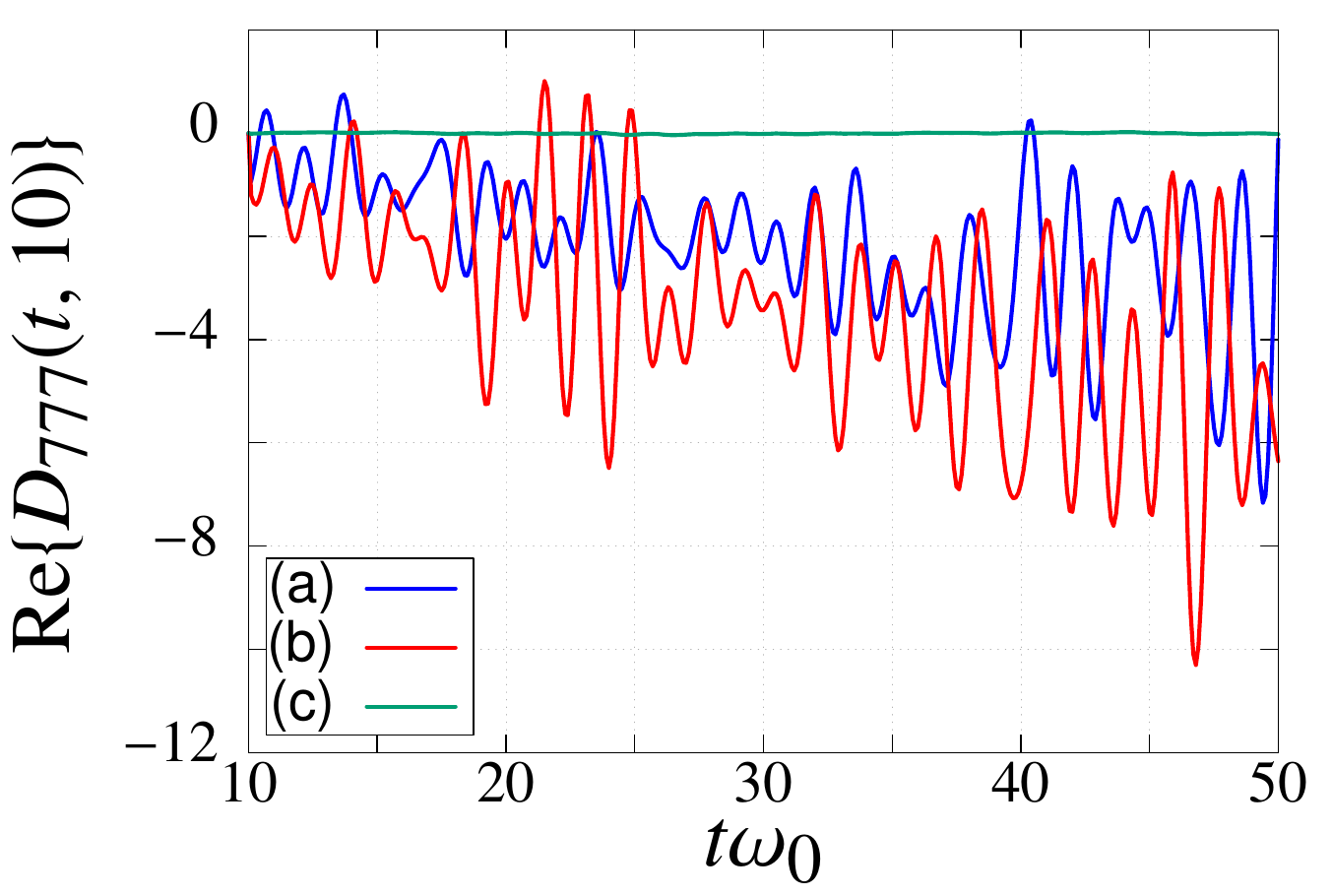}
\caption{
Real part of the three-body correlation function $D_{777}(t, t')$ for the central spin in a 1D $XXZ$ spin lattice in arbitrary units. The time $t'$ is fixed as $t' = 10$.  The blue, red, and green curves are for (a) a TLS + spin lattice + QT, (b) a TLS + spin lattice, and (c) an isolated spin lattice.
The curve (a) is replotted from Fig.~\ref{fig:3corr}.
We cannot define the temperature for (b) and (c) because those systems do not couple to the QT.
\label{fig:3corrWO}
}
\end{figure}
  
Figure~\ref{fig:3corrWO} displays the real part of the three-body correlation function $D_{777}(t, 10)$ for (a) a TLS + spin lattice + QT ($\epsilon/\omega_0 = 1$, $\hbar \zeta = 0.01$, $\beta \hbar \omega_0 \to \infty$, blue), (b) a TLS + spin lattice  ($\epsilon/\omega_0 = 1$, $\hbar \zeta = 0$, red), and (c) an isolated spin lattice ($\epsilon/\omega_0 = 0$, $\hbar \zeta = 0$, green).  The parameters $\epsilon_0$, $\zeta$, and $\beta$ are the same as in Fig.~\ref{fig:2corrWO}.
For the isolated spin lattice (c), the three-body correlation function is almost zero.
This arises from the isotropy of the $XXZ$ spin lattice.
By introducing TLS--spin lattice coupling ($\epsilon_0 / \omega_0 \neq 0$), the isotropy is broken and the amplitude of the three-body correlation function increases in time.
This indicates that the TSL contributes to the properties of the three-body correlation function $D_{777}(t, t')$ rather than the spin-lattice system. This tendency is opposite to that for a two-body correlation function. The oscillatory frequencies of the two-body and three-body correlation functions are also different. 

\section*{Data availability}
The data that support the findings of this study are available from the corresponding author upon reasonable request.

\bibliography{reference,referenceSB,tanimura_publist}

\end{document}